\documentclass[12pt]{iopart}
\pdfoutput=1
\usepackage{iopams}
\expandafter\let\csname equation*\endcsname\relax

\expandafter\let\csname endequation*\endcsname\relax
\usepackage{amsmath}
\usepackage{graphicx}
\usepackage{hyperref}
\usepackage{cite}
\usepackage{braket}
\usepackage{color}
\usepackage{comment}
\usepackage{ulem}

\newcommand{\lac}[1]{{\color{red}#1}}

\begin{document}
\title[Ultracold molecules for quantum simulation]{Ultracold molecules for quantum simulation: rotational coherences in CaF and RbCs}
	
\author{Jacob~A~Blackmore$^1$\footnotemark, 
Luke Caldwell$^{2}$\footnotemark[\value{footnote}], 
Philip~D~Gregory$^1$, 
Elizabeth~M~Bridge$^1$, 
Rahul Sawant$^1$, 
Jes{\'u}s~Aldegunde$^{3}$, 
Jordi~Mur-Petit$^4$, 
Dieter~Jaksch$^{4,5}$, 
Jeremy~M~Hutson$^{6}$,
B~E~Sauer$^{2}$,
M~R~Tarbutt$^{2}$ 
and Simon~L~Cornish$^1$}
\address{$^1$Joint Quantum Centre (JQC) Durham-Newcastle, Department of Physics, Durham University, South Road, Durham DH1 3LE, United Kingdom.\\
		$^{2}$ Centre for Cold Matter, Blackett Laboratory, Imperial College London, Prince Consort Road, London SW7 2AZ, United Kingdom.\\
		$^3$ Departamento de Quimica Fisica, Universidad de Salamanca, 37008 Salamanca, Spain. \\
		$^4$ Clarendon Laboratory, University of Oxford, Parks Rd, Oxford OX1 3PU, United Kingdom.\\
        $^{5}$ Centre for Quantum Technologies, National University of Singapore, 3 Science Drive 2, 117543 Singapore \\
		$^{6}$ Joint Quantum Centre (JQC) Durham-Newcastle, Department of Chemistry, Durham University, South Road, Durham DH1 3LE, United Kingdom.\\
		}
 \footnotetext{These authors contributed equally to this work.}      \eads{\mailto{m.tarbutt@imperial.ac.uk}, \mailto{s.l.cornish@durham.ac.uk}}
	
\abstract{
Polar molecules offer a new platform for quantum simulation of systems with long-range interactions, based on the electrostatic interaction between their electric dipole moments.
Here, we report the development of coherent quantum state control using microwave fields in $^{40}$Ca$^{19}$F and $^{87}$Rb$^{133}$Cs molecules, a crucial ingredient for many quantum simulation applications. We perform Ramsey interferometry measurements with fringe spacings of $\sim 1~\rm kHz$ and investigate the dephasing time of a superposition of $N=0$ and $N=1$ rotational states when the molecules are confined. For both molecules, we show that a judicious choice of molecular hyperfine states minimises the impact of spatially varying transition-frequency shifts across the trap. For magnetically trapped $^{40}$Ca$^{19}$F we use a magnetically insensitive transition and observe a coherence time of 0.61(3)~ms. For optically trapped $^{87}$Rb$^{133}$Cs we exploit an avoided crossing in the AC Stark shifts and observe a maximum coherence time of 0.75(6)~ms. }


\maketitle

\end{indented}


\noindent
Since Lloyd's proof that evolving a controllable quantum system in small time steps can allow efficient simulation of any quantum many-body Hamiltonian~\cite{Lloyd1996}, the field of quantum simulation has grown enormously. This growth has been driven by the prospect of understanding complex physical systems, such as high-temperature superconductors~\cite{Manousakis:2002,Byrnes2008} or warm, dense nuclear matter~\cite{Zohar2016}. In this context, the role of the quantum simulator is to implement a model of a physical system, and to measure observables that can be compared to real systems of interest~\cite{Johnson2014}. Currently, a number of experimental platforms are under exploration, each in essence attempting to address Cirac and Zoller's criteria to qualify as a quantum simulator~\cite{Cirac2012} through different paths~\cite{Georgescu2014,Glaetzle:2015,Glaetzle:2017}.

Trapped ions are the best-established candidate system for quantum simulation~\cite{Blatt:2012}. They offer exceptional quantum control over small numbers of ions with high-fidelity gate operations~\cite{Leibfried2003rmp}. Ongoing efforts are aimed at scaling up to larger numbers of ions. In the short term, very good control is probably achievable for 10 to 20 ions~\cite{Friis:2018}, and less perfect control for up to 50 ions. Scaling up to much larger numbers is a challenging goal that will probably require new ideas. In contrast, superconducting circuits have recently made considerable progress as an alternative platform~\cite{Houck:2012}. Here, the superconducting qubits and qubit gates are of lower fidelity, but scaling up to larger systems is easier and integration with other existing electronic technology is relatively straightforward.

Quantum simulation with ultracold atoms in optical lattices~\cite{Gross:2017,LewensBook} has offered some ground-breaking results, such as the recent observation of magnetic correlations in experiments implementing the two-dimensional Hubbard model~\cite{Parsons2016,Boll2016,Cheuk2016}. These experiments prepare a large number of atoms in a well-controlled initial state. Gate operations are highly parallelisable, so this approach is suitable for quantum simulation of model Hamiltonians that do not require spatially varying operations. Indeed, it has frequently been difficult to address individual atoms, although the development of the quantum gas microscope \cite{Bakr:2009,Sherson:2010,Cheuk:2015} and tweezer arrays \cite{Lester:2015,Endres:2016,Barredo:2016} represent important breakthroughs in this respect. Gate operations are relatively slow, governed by trap frequencies and tunnelling rates on the order of Hz to kHz~\cite{Jaksch:1999}. Moreover, interactions between ground-state alkali-metal atoms are short-ranged, which makes them suitable to model local interactions only. This last point has prompted growing interest in long-range dipolar interactions, which may be implemented in atomic systems either by using highly magnetic atoms~\cite{Griesmaier:2005,Lu:2011,Aikawa:2012,Baier:2016,Wenzel:2017,Chomaz:2018} or by exciting atoms to Rydberg states~\cite{Jaksch:2000,urban2009,Gaetan:2009,Glaetzle:2017}.

Ultracold polar molecules offer new possibilities for quantum simulation. The electric dipole moments of polar molecules give rise to interactions that are significantly greater than those between magnetic atoms. Although interactions between Rydberg states are even stronger, ground-state polar molecules have longer lifetimes. Molecules also possess a rich internal structure, with vibrational and rotational degrees of freedom, in addition to electronic and hyperfine structure. The rotational structure is of particular importance in the context of quantum simulation, providing a rich basis in which to encode pseudo-spins that can be easily manipulated with microwave fields. Moreover, the electric dipole coupling between rotational states allows tunable long-range interactions to be engineered between the encoded spins. These properties have inspired numerous proposed applications in quantum magnetism, the study of the many-body physics of coupled spins
\cite{Barnett:2006,Micheli:2006,Gorshkov:2011,Gorshkov2011PRA,Zhou:2011,Hazzard:2013,Capogrosso-Sansone:2010,Pollet:2010,Lechner:2013}.
Finally, the enormous range of molecular species allows selection of molecular properties to match the application. For example, molecules with no electronic spin or orbital angular momentum possess only an electric dipole moment, whereas those with an unpaired spin may have both electric and magnetic dipole moments. Beyond the field of quantum simulation, ultracold molecules also have potential applications in the study of quantum-controlled chemistry \cite{Krems2008,Ospelkaus:2010Science,Miranda2011,Ye2018}, quantum information processing \cite{DeMille2002} and precision measurement \cite{Flambaum2007,Isaev2010,Hudson2011,Baron2013} .

There has been considerable success in producing a growing number of ultracold molecular species, both through the association of atoms in ultracold atomic gases and more recently by direct laser cooling of molecules. The most successful association method to date has employed magnetoassociation on a zero-energy Feshbach resonance~\cite{Koehler2006,Chin2010} followed by optical transfer using  stimulated Raman adiabatic passage (STIRAP)~\cite{Bergmann1998} to produce molecules in the rovibronic ground state. This approach has been employed extensively to associate pairs of ultracold alkali-metal atoms, where the ability to cool the atomic gases to quantum degeneracy leads to molecular gases at high phase-space density and typical temperatures of $\sim 1~\mu$K or below. Ultracold ground-state KRb~\cite{Ni2008}, Cs$_2$~\cite{Danzl:2008}, Rb$_2$~\cite{Lang:2008}, RbCs~\cite{Takekoshi:2014,Molony:2014}, NaK~\cite{Park:2015,Seeselberg2018}, NaRb~\cite{Guo:2016} and NaLi~\cite{Rvachov:2017} molecules have all been created. A new wave of experiments aims to extend this approach to molecules with doublet ground states, by associating atoms in mixtures of alkali-metal and closed-shell atoms~\cite{Munchow:2011, Barbe:2018, Guttridge2018}. 
A new technique that is advancing very rapidly is direct laser cooling of molecules. Although their complex level structure makes molecules difficult to cool, there are many molecules that have almost-closed electronic transitions suitable for laser cooling. So far, laser cooling has been demonstrated for SrF~\cite{Shuman2010,Barry:2014,McCarron2015,Norrgard2016}, YO~\cite{Hummon2013}, CaF~\cite{Zhelyazkova2014,Truppe2017,Anderegg2017}, YbF~\cite{Lim:2018} and SrOH~\cite{Kozyryev2017}. Cooling of several other species is also being pursued, including BaF~\cite{Chen2017}, BaH~\cite{Iwata2017} and TlF~\cite{Hunter2012}. Recent advances have led to laser cooling of molecules to sub-Doppler temperatures~\cite{Truppe2017}, and transfer of these ultracold molecules into magnetic and optical traps~\cite{Williams:2018,McCarron:2018,Anderegg2018}. Direct cooling into the microkelvin regime has also been demonstrated by optoelectrical Sisyphus cooling~\cite{Prehn2016}. With such remarkable and rapid progress, we can expect the control of ultracold molecules soon to develop to the level required for proof-of-principle quantum simulation.

In this paper we explore the potential of ultracold molecules for quantum simulation. We focus on two contrasting molecular species, $^{40}$Ca$^{19}$F and $^{87}$Rb$^{133}$Cs, hereafter referred to simply as CaF and RbCs.
In section~\ref{Sec:QSim} we describe the properties of molecules that make them suited to applications in quantum simulation, focussing on the role of the molecular rotation. In section \ref{sec:CaFstatecontrol} we describe the rotational and hyperfine structure of CaF and RbCs, which are important for internal state control using microwave fields.
In section~\ref{Sec:Ramsey} we report Ramsey interferometry measurements and investigate the coherence time of a superposition of ground and excited rotational states when the molecules are confined. For both molecules, we show that a judicious choice of molecular hyperfine states minimises the impact of spatially varying shifts in transition frequencies across the trap, providing a route to long interrogation and control times in future experiments. Finally, we outline our plans for loading, addressing and detecting individual molecules in ordered arrays; this is the next major challenge in the development of molecules for quantum simulation.

\section{Quantum simulation with ultracold molecules\label{Sec:QSim}}
\begin{figure}
\includegraphics[width=\textwidth]{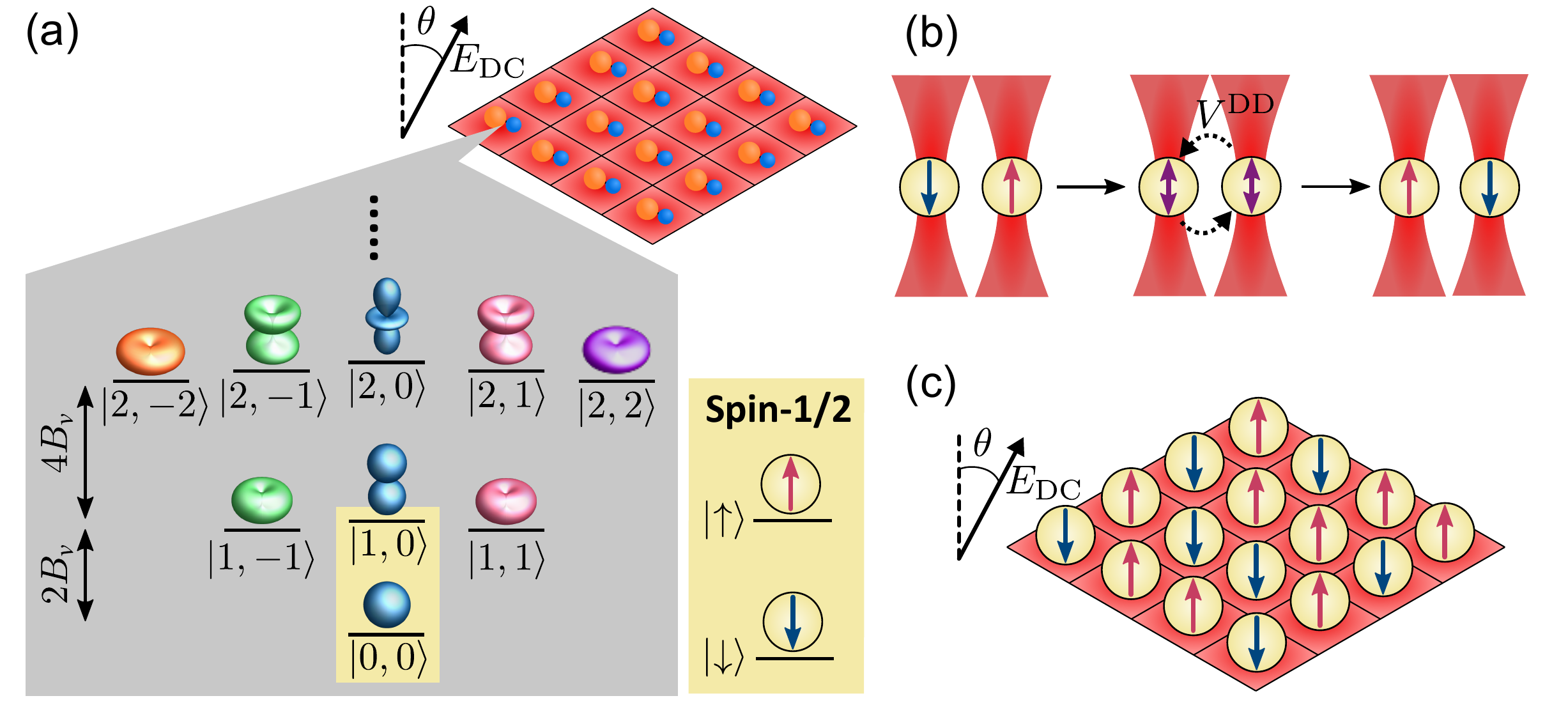}
\caption{
Using polar molecules for quantum simulation. (a) Pseudo-spins can be encoded in the internal rotational states of molecules (blue-orange dumbbells) confined in regular arrays. In this example, we map the two spin states $\ket{\downarrow}$, $\ket{\uparrow}$ onto the rotational states $\ket{N=0,M_N=0}$ and $\ket{1,0}$. The energy separation between these states is set by the rotational constant $B_v$ and lies in the microwave domain. (b) Dipole-dipole interactions ($V^\mathrm{DD}$) lead to spin-exchange (or spin flip-flop) interactions between adjacent molecules, here shown confined in individual tightly-focussed optical tweezers. (c) In a deep optical lattice, the molecules can be used to simulate models of quantum magnetism, such as the XXZ model described in the text. The applied static electric field can be used to tune the model parameters.}
\label{Fig:Summary}
\end{figure}

In this section we introduce the distinctive  properties of polar molecules that make them an attractive platform for quantum simulation. We motivate our study of rotational coherence by illustrating how the rotational degrees of freedom can be used to encode models of quantum magnetism. We limit our discussion to the simple case of diatomic polar molecules in their electronic ground state, pinned on the sites of an optical lattice or tweezer array such that motion and tunnelling between sites can be neglected. Extensions beyond this simple scenario to include, for example, different trapping geometries, tunnelling between lattice sites, disordered filling, and greater molecular complexity, lead to even richer physics~\cite{Baranov:2012,Gorshkov2013,Wall:2015} but are beyond the scope of this paper.

Diatomic molecules feature two new degrees of freedom compared to atoms: \textit{vibrations}, corresponding to the variation of the distance, $R$, between the constituent nuclei, and characterised by a vibrational quantum number $v=0,1,2,\ldots$; and \textit{rotations} of the molecule about the axis perpendicular to the internuclear separation vector. These are described by a rotational quantum number, $N=0,1,2,\ldots$, and its projection on the laboratory quantisation axis, $M_N$~\cite{BransdenBook}. The typical energy scale associated with vibrations is 50~K or more; the ultracold molecules we consider are in a single vibrational state, and we focus on the rotational degree of freedom. The energies of the rotational states are approximately $E_N=B_vN(N+1)$, where $B_v$ is the rotational constant. Typically, in the vibrational ground state $B_0 \geq h \times 400~\mathrm{MHz} \simeq k_{\rm{B}} \times 20 ~\mathrm{mK}$ and the rotational states can be conveniently manipulated with microwave fields.  Molecules also possess rich hyperfine structure, as discussed in section~\ref{sec:CaFstatecontrol}, but this complication does not preclude the use of rotational states for quantum simulation.

It is possible to encode pseudo-spins in molecular rotational states in order to realise various models of quantum magnetism. For example, Barnett \textit{et al.}~\cite{Barnett:2006} proposed modelling an effective spin-1/2 particle by identifying the state $\ket{\downarrow}$ with the rotational ground state, $\ket{N=0, M_N=0}$ of a $^1\Sigma$ molecule, and the state $\ket{\uparrow}$ with one of the components of the first rotationally excited state, $\ket{N=1, M_N=-1,0,+1}$, as illustrated in figure~\ref{Fig:Summary}\,(a).
The motivation for using molecules for quantum simulation stems from the ability to engineer controllable long-range interactions between spins via the electric dipole-dipole interaction (DDI) between molecules $i$ and $j$, 

\begin{align}
V_{ij}^{\mathrm{DD}}
 &=
 \frac{1}{4\pi\epsilon_0}
 \frac{ \vec{\mu}_i \cdot \vec{\mu}_j
       - 3(\vec{\mu}_i \cdot\vec{e}_{ij}) (\vec{\mu}_j \cdot\vec{e}_{ij}) }{r_{ij}^3} \:.
 \label{Eq:DDI}
\end{align}
Here molecule $i$ has electric dipole moment $\vec{\mu}_i$ and position vector $\vec{r}_i$; $\vec{r}_{ij}=\vec{r}_i-\vec{r}_j$, and $\vec{e}_{ij}$ is the unit vector in the direction of $\vec{r}_{ij}$. It is important to recognise that, in the absence of fields breaking rotational symmetry, molecular eigenstates have a vanishing electric dipole moment. 
However, it is possible to produce \textit{controllable} long-range anisotropic DDI by using (i) an external DC electric field to mix rotational states with the same $M_N$ and orient the dipole moments in space, and/or (ii) AC fields to create a superposition of $\ket{N,M_N}$ states with an oscillating electric dipole moment~\cite{Buchler:2007}.
Typical achievable dipole moments are $\sim 1$ Debye, leading to a DDI energy $\sim h\times 1$~kHz between neighbouring molecules spaced by $532$\,nm in an optical lattice. This energy scale sets the coherence time needed for applications in quantum simulation.

In the space spanned by the internal states $\ket{\uparrow},\ket{\downarrow}$, the effect of the DDI is to couple the two-molecule states $\ket{\uparrow\downarrow}$ and $\ket{\downarrow\uparrow}$,
\begin{align}
 V^{\mathrm{DD}}\ket{\uparrow\downarrow}
 = a \ket{\uparrow\downarrow}
 + b \ket{\downarrow\uparrow},
 \label{eq:flip-flop}
\end{align}
with $a = \braket{\uparrow\downarrow|V^\mathrm{DD}|\uparrow\downarrow}$
and  $b = \braket{\downarrow\uparrow|V^\mathrm{DD}|\uparrow\downarrow}$; here, $\ket{s_1 \,s_2}$ indicates that the first molecule is in internal state $s_1$ and the second in $s_2$, with $s_j\in\{\uparrow,\downarrow\}$.
In the language of spins, the dynamics triggered by the off-diagonal element $b$ is a flip-flop, or transfer of the excitation between the two molecules.
In this language, the DDI between two molecules located in sites $i,j$ may be written
\begin{align}
 V_{ij}^{\mathrm{DD}}
 = \frac{J_\perp}{2} \left(S_i^+ S_j^- + \rm{H.c.}\right) + J_z S_i^z S_j^z \:.
 \label{eq:Vddi}
\end{align}
Here
$S_i^z=\frac{1}{2}\left( \ket{\uparrow_i}\bra{\uparrow_i} - \ket{\downarrow_i}\bra{\downarrow_i} \right)$, 
$S_i^+ = \ket{\uparrow_i}\bra{\downarrow_i}$ and $S_i^- = (S_i^+)^\dagger$ are the spin operators for the molecule at site $i$, and the parameters $J_{\perp,z}$ depend on the magnitude and orientation of the electric dipole moment of the molecules in states $\ket{\uparrow},\ket{\downarrow}$.

For a collection of molecules pinned to the sites of a two-dimensional (2D) lattice in the presence of an applied DC electric field, this two-body interaction leads to the well-known XXZ spin Hamiltonian~\cite{Gorshkov:2011,Gorshkov2011PRA}
\begin{align}
 H_\mathrm{XXZ}
 &= \sum_{i\neq j}\left[ \frac{J_\perp}{2} \left(S_i^+ S_j^- + \rm{H.c.}\right)
  + J_z S_i^z S_j^z \right] \:.
 \label{eq:XXZ}
\end{align}
We choose to map
$\ket{\downarrow},\ket{\uparrow}$ onto the states that adiabatically correlate with $\ket{N=0,M_N=0}$ and $\ket{N=1,M_N=0}$ at zero field, see figure~\ref{Fig:Summary}(a).
With this choice, the couplings are
\begin{align}
 J_\perp
 &=\frac{1-3\cos^2\theta_{ij}}{4\pi\epsilon_0 
 r_{ij}^3}\mu_{\uparrow\downarrow}^2 \:,
 &
 J_z
 &=\frac{1-3\cos^2\theta_{ij}}{4\pi\epsilon_0 r_{ij}^3}
 \left( \mu_{\uparrow\uparrow} - \mu_{\downarrow\downarrow}
 \right)^2,
 \label{eq:Js}
\end{align}
with $\mu_{s_is_j}=\braket{s_i|\hat{\mu}_0|s_j}$ the elements of the electric dipole operator in the direction of the applied DC electric field, and $\theta_{ij}$ the angle between the molecular dipole moments and the intermolecular vector~\cite{Gorshkov:2011,Gorshkov2011PRA,Wall:2015}.
Eq.~\eqref{eq:Js} illustrates the tunability offered by molecules for quantum simulation: even in this simple example, the strength and even the sign of the spin-spin couplings can be controlled in a dynamical way by means of external fields.

The spin dynamics brought about by the DDI described by the first term in equation~\eqref{eq:Vddi} were first observed in a collection of polar molecules by Yan et al.~\cite{Yan2013}.
In this experiment, $^{40}$K$^{87}$Rb molecules in their electronic and vibrational ground state were trapped in a three-dimensional optical lattice in the absence of an applied electric field. The molecules were initialized in state $\ket{\downarrow}=\ket{0,0}$ and, after a Ramsey sequence of variable duration, the number of molecules remaining in $\ket{\downarrow}$ was measured. The observations revealed an oscillatory behaviour on top of an overall decay of coherence, with a dominant frequency $\approx48$~Hz, which is close to the strength of the DDI with nearest-neighbour molecules 532~nm away, $J_{\perp}/2\simeq h \times 52$~Hz~\cite{Yan2013}. 

This simple example highlights the potential of ultracold molecules in the quantum simulation of many-body Hamiltonians with long-range interactions. These include quantum spin models 
\cite{Micheli:2006,Barnett:2006,Gorshkov:2011,Gorshkov2011PRA,Zhou:2011,Hazzard:2013}, novel phases of quantum matter such as super-solids~\cite{Capogrosso-Sansone:2010,Pollet:2010}, spin glasses~\cite{Lechner:2013} and phases with topological order~\cite{Buchler:2007,Manmana:2013}; we refer the reader to the recent reviews~\cite{Baranov:2012,Gorshkov2013,Wall:2015} for further details.
For the realisation of these proposals, many experimental challenges remain to be addressed, the most salient ones being (i) full site-resolved coherent control and measurement of the internal states, (ii) long coherence times allowing time-dependent studies, and (iii) large filling fractions of the lattice.
In the following we demonstrate rotational-state Ramsey interferometry with ultracold ground-state RbCs and CaF molecules as the archetypal example of \lac{a} coherent control technique\sout{s}, and show how to extend the coherence times, thus advancing the prospects for quantum simulation with molecules.

\section{Energy levels in electric and magnetic fields}
\label{sec:CaFstatecontrol}

Quantum control of the internal state of a molecule requires a good understanding of the hyperfine structure and the Zeeman and Stark effects. An effective Hamiltonian describing the rotational and hyperfine structure in the ground electronic state is 
\begin{equation}
H = H_{\rm rot} + H_{\rm hyp} + H_{\rm Zeeman} + H_{\rm DC~Stark}+ H_{\rm AC~Stark}.
\label{Eq:HamCaF}
\end{equation}
The rotational part is
\begin{equation}
H_{\rm rot} = B_v \vec{N}^{2} - D_v \vec{N}^{4},
\label{Eq:HamRot}
\end{equation}
where $\vec{N}$ is the dimensionless rotational angular momentum operator. The most precise values for the rotational constants $B_0$ and $D_0$ are given in \cite{Anderson1994} for CaF and in \cite{Gregory:2016} for RbCs. The DC Stark Hamiltonian is adequately modelled using a rigid-rotor model,
\begin{equation}
H_{\rm DC~Stark} = -\mu_\mathrm{e} \vec{E}\cdot\hat{z},
\label{Eq:HamStark}
\end{equation}
where $\vec{E}$ is the applied electric field, $\mu_\mathrm{e}$ is the electric dipole moment in the molecule-fixed frame, and $\hat{z}$ is a unit vector in the direction of the internuclear axis. For CaF, $\mu_\mathrm{e}= 3.07(7)$~D~\cite{Childs1984}, while for RbCs, $\mu_\mathrm{e} = 1.225(3)(8)$~D~\cite{Molony:2014}. 
The AC Stark interaction of the molecule with light of intensity $I$ is
\begin{equation}
H_{\text{AC Stark}}=-\frac{1}{2\epsilon_0 c}\alpha I,
\label{eq:AC Stark}
\end{equation}
where $\alpha$ is the frequency-dependent polarisability tensor. This interaction is responsible for the trapping potential in an optical trap, and also affects the energy difference between states. For linearly polarized light, we can write the polarisability as 
\begin{equation}
\label{eq:polarisability}
\alpha(\theta) = \alpha^{(0)}+\alpha^{(2)}P_2\left(\cos \theta \right),
\end{equation}
where $\alpha^{(0)}$ and $\alpha^{(2)}$ are the spatially isotropic and anisotropic components of the polarisability tensor and $\theta$ is the angle between the internuclear axis of the molecule and the polarisation of the light. The isotropic component of the polarisability affects all rotational states in the same way, contributing only to the trapping potential. In contrast, because of the $\theta$ dependence, the anisotropic component depends on both $N$ and $M_N$ and directly affects the transition frequency between rotational states. 

The hyperfine and Zeeman parts of the Hamiltonian \cite{Aldegunde:2008,Aldegunde:2018} vary between the two molecular species we consider due to their different electronic structures. In the following, we describe these parts separately for CaF and RbCs.

\begin{figure}[t!]
	\centering
	\includegraphics[width=\textwidth]{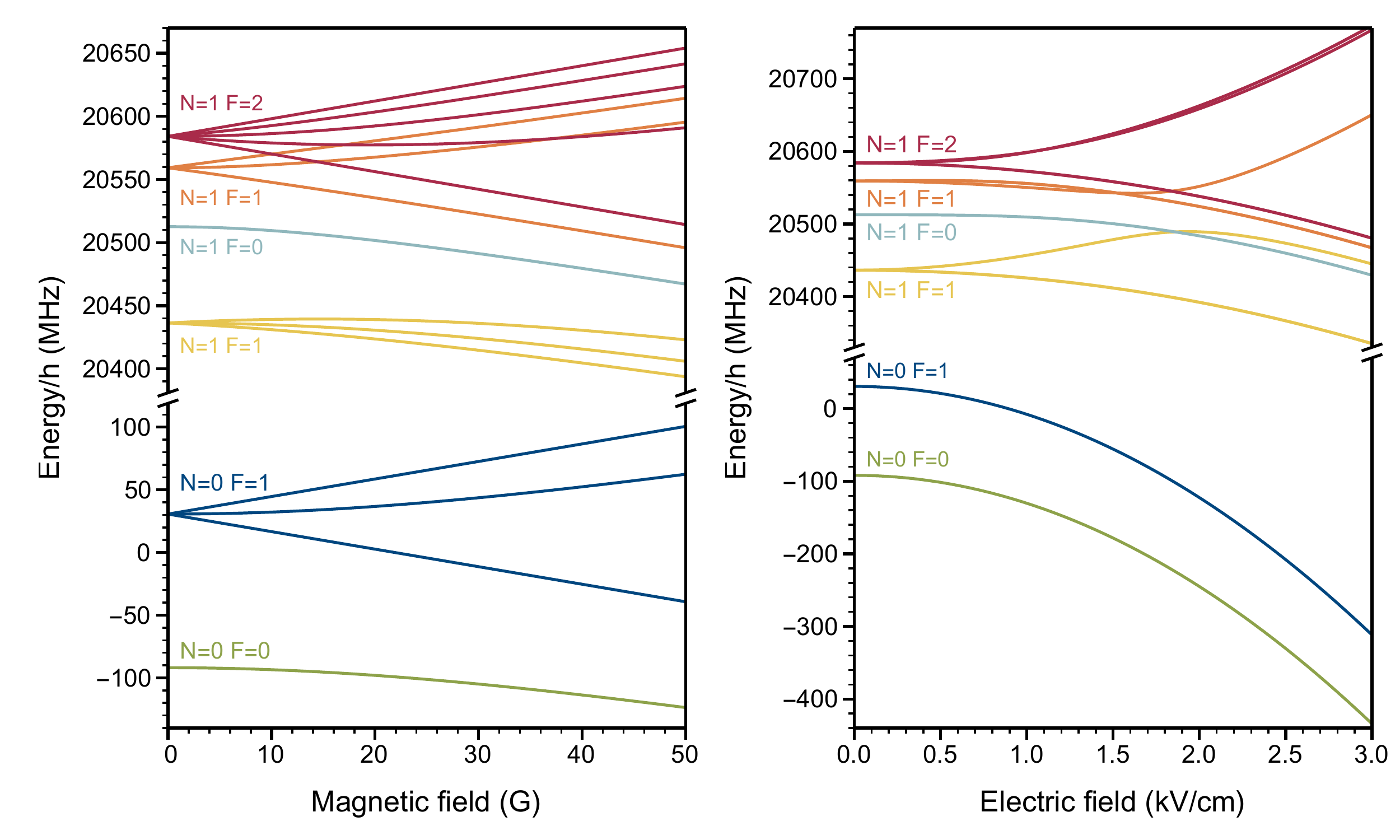}
	\caption[]{The $N=0$ and $N=1$ levels of the X$^{2}\Sigma^{+} (v=0)$ state of CaF. (a) Energies as a function of magnetic field, at zero electric field. (b) Energies as a function of electric field, at zero magnetic field. Note the breaks in the vertical axes.}
	\label{fig:CaF_Zeeman_Stark}
\end{figure}

\subsection{CaF}
\label{Sec:Structure_CaF}
For CaF, which has one unpaired electron, the hyperfine component of the Hamiltonian is~\cite{Childs1981,Brown:2003}
\begin{equation}
H_{\rm hyp}^{\mathrm{CaF}} =\gamma \vec{S}\cdot\vec{N} + (b+c/3) \vec{I}_\mathrm{F}\cdot\vec{S} + (c/3) \sqrt{6}T^2(C)\cdot T^2(\vec{I}_\mathrm{F},\vec{S}) + c_{\mathrm{F}} \vec{I}_\mathrm{F}\cdot\vec{N},
\label{Eq:HamHyp}
\end{equation}
where we have introduced the dimensionless operators for the electron spin, $\vec{S}$, and the fluorine nuclear spin, $\vec{I}_\mathrm{F}$. The $^{40}$Ca isotope has no nuclear spin. The first term in equation~\eref{Eq:HamHyp} is the electron spin-rotation interaction. The second and third terms account for the interaction between the electron and nuclear magnetic moments, written here in spherical tensor form; $T^2(\vec{I}_\mathrm{F},\vec{S})$ denotes the rank-2 spherical tensor formed from $\vec{I}_\mathrm{F}$ and $\vec{S}$, while $T^2(C)$ is a spherical tensor whose components are the (renormalised) spherical harmonics $C^2_q(\theta,\phi)$. The spectroscopic parameters $b$ and $c$ are those of Frosch and Foley \cite{frosch:pr1952}, which are sometimes collected together as a Fermi contact parameter $b_{{\rm F}}=b+c/3$ and a dipolar parameter $t=c/3$. The last term is the nuclear spin-rotation interaction, and is three orders of magnitude smaller than the others. Precise values for $\gamma$, $b$, $c$ and  $c_{\mathrm{F}}$ (sometimes called $C$) are given in~\cite{Childs1981}.

The effective Zeeman Hamiltonian is \cite{Brown:2003}
\begin{equation}
H_{\rm Zeeman}^\mathrm{CaF} = g_{S} \mu_\mathrm{B} \vec{S}\cdot\vec{B} + g_{l} \mu_\mathrm{B} \left[\vec{S}\cdot\vec{B}-(\vec{S}\cdot\hat{z})(\vec{B}\cdot\hat{z})\right] - g_{\rm r}\mu_\mathrm{B} \vec{N}\cdot\vec{B} - g_\mathrm{N}^\mathrm{F}\mu_\mathrm{N}\vec{I}_\mathrm{F}\cdot\vec{B},
\label{Eq:HamZeeman}
\end{equation}
where $\vec{B}$ is the applied magnetic field. The terms represent, from left to right, the electronic Zeeman interaction characterised by $g_S$, its anisotropic correction characterised by $g_l$, the rotational Zeeman interaction characterised by $g_\mathrm{r}$, and the nuclear Zeeman interaction characterised by the nuclear $g$-factor $g_{\rm N}^{\rm F}$, which is defined to include the small effects of diamagnetic shielding. The last three terms are typically three orders of magnitude smaller than the first term, but are important when searching for magnetically insensitive transitions. The value of $g_l$ can be estimated using Curl's approximation  $g_l \approx -\gamma/2B$~\cite{curl:mp1965,Devlin2015}.


Figure~\ref{fig:CaF_Zeeman_Stark} shows the eigenvalues of \eref{Eq:HamCaF} corresponding to the ground and first-excited rotational levels, $N=0$ and $N=1$, as functions of magnetic and electric fields. At low field, the states are best described in the coupled representation $|N,S,I_\mathrm{F},F,M_{F}\rangle$, while at high field they are best described in the uncoupled representation $|N,M_{N}\rangle|S,M_{S}\rangle|I_\mathrm{F},M_{I_\mathrm{F}}\rangle$. The Zeeman shifts shown in figure~\ref{fig:CaF_Zeeman_Stark}(a) are relevant for trapping molecules in the MOT and in a magnetic trap. At low magnetic field, $B \lesssim 5$~G, these shifts are well approximated as $\Delta E_{\rm Zeeman} = g_{F} \mu_\mathrm{B} B M_{F}$. For the six zero-field energy levels shown in the figure, in order of increasing energy, the values of $g_{F}$ are 0, 1.000, $-0.294$, 0, 0.795 and 0.500. The Stark shifts shown in figure~\ref{fig:CaF_Zeeman_Stark}(b) are relevant for trapping molecules electrically, and for inducing the electric dipole moments needed for quantum simulation. The Stark shifts are quadratic at low field and become linear at sufficiently high field. The dipole moment is the gradient of the Stark shift and saturates slowly towards $\mu_\mathrm{e}$. For $N=0$, the dipole moment is 1~D when $E = 7.5$~kV\,cm$^{-1}$, and increases to 2.4~D when $E=75$~kV\,cm$^{-1}$.

\subsection{RbCs}
\label{sec:RbCs MW Control}

\begin{figure}[t!]
	\centering
	\includegraphics[width=\textwidth]{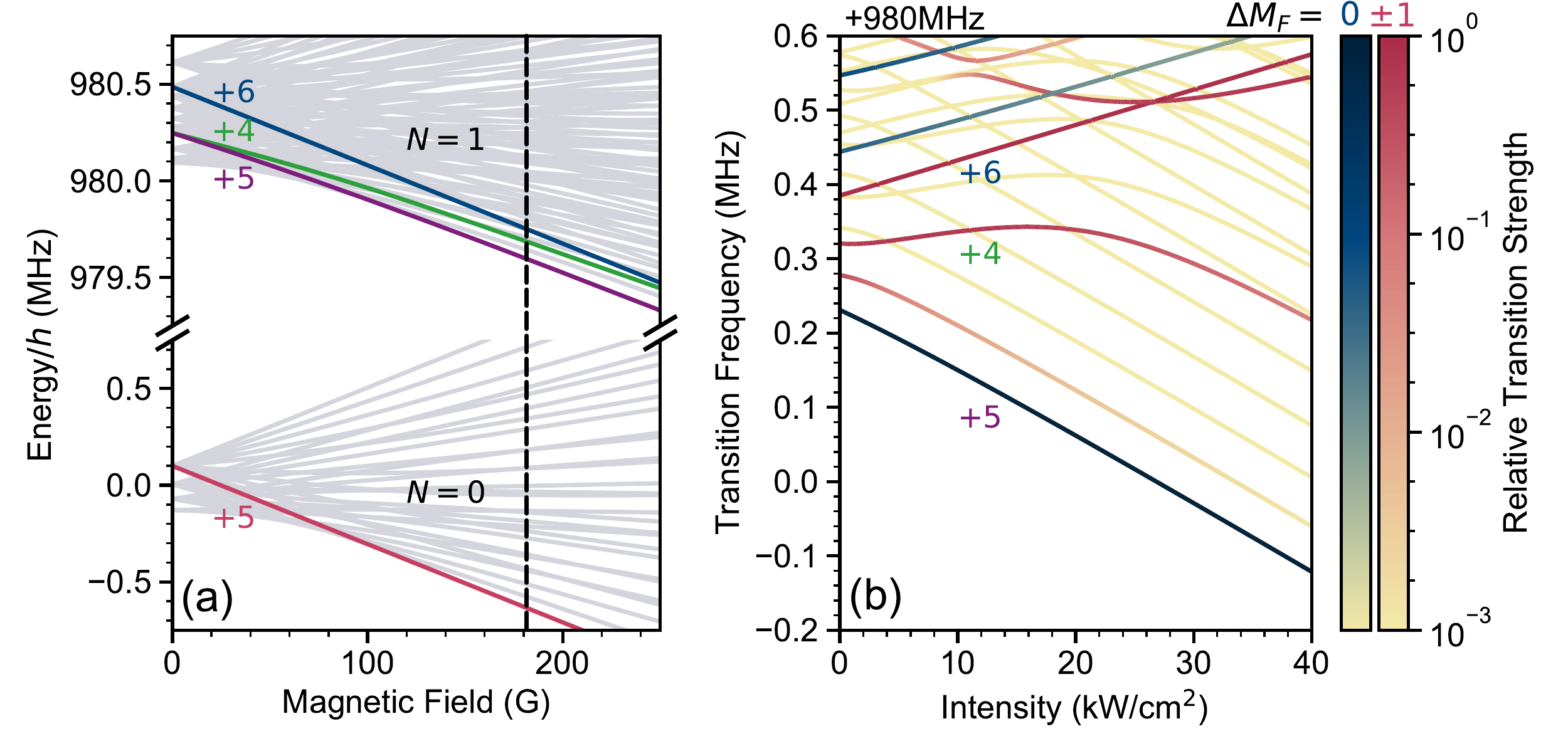}
	\caption[]{The hyperfine structure of the $N=0$ and $N=1$ levels of the $X^1\Sigma^+ (v=0)$ state of RbCs. (a) Energies as a function of magnetic field with the states relevant to this work highlighted and labelled by $M_F$. STIRAP populates $\ket{N=0,M_F=+5}$ at a magnetic field of 181.5~G (indicated by the vertical dashed line). (b) Transition frequencies as a function of laser intensity for microwave transitions from $\ket{N=0,M_F=+5}$ to states in $N=1$ in the presence of light at a wavelength of $1550~\mathrm{nm}$ polarised parallel to an applied magnetic field of 181.5~G. Transitions relevant to this work are labelled by $M_F$ of the upper state. The colour indicates the relative strengths for $\Delta M_F=0$ (blue) and $\Delta M_F=\pm 1$ (red) transitions.}
	\label{fig:RbCs Structure}
\end{figure}

In RbCs there are no unpaired electrons, so the hyperfine Hamiltonian is dominated by the nuclear interactions~\cite{Ramsey:1952,Brown:2003,Bryce:2003,Aldegunde:2008} 
\begin{equation}
H_{\rm hyp}^{\mathrm{RbCs}} = \sum_{i=\mathrm{Rb,Cs}}e{\mathbf Q}_\mathrm{i}\cdot\mathbf{q}_\mathrm{i} + \sum_{i=\mathrm{Rb,Cs}}c_i\vec{I}_i\cdot\vec{N} - c_3  \sqrt{6}T^2(C)\cdot T^2(\vec{I}_\mathrm{Cs},\vec{I}_\mathrm{Rb}) + c_4\vec{I}_\mathrm{Cs}\cdot\vec{I}_\mathrm{Rb}.
\label{eq:Hyp_RbCs}
\end{equation}
Here, the values of the component nuclear spins are  $I_{\mathrm{Rb}} = 3/2$ and $I_{\mathrm{Cs}}=7/2$. The first term is the electric quadrupole interaction  and represents the interaction between the nuclear electric quadrupole of nucleus i ($e{\mathbf Q}_\mathrm{i}$) and the electric field gradient at the nucleus ($\mathbf{q}_\mathrm{i}$). This term exists only for nuclei with $I\ge1$, so is absent in CaF; its strength is proportional to the coupling constants $(eQq)_{{\rm Rb}}$ and $(eQq)_{{\rm Cs}}$. The second term is the interaction between the nuclear magnetic moments and the magnetic field created by the rotation of the molecule, with spin-rotation coupling constants $c_\mathrm{Rb}$ and $c_\mathrm{Cs}$. The two remaining terms represent the tensor and scalar interactions between the nuclear dipole moments, with spin-spin coupling constants $c_3$ and $c_4$, respectively. The quantity $c_3$ has both direct dipolar and indirect (electron-mediated) contributions, while $c_4$ arises entirely from indirect  interactions. The values for the relevant coefficients are given in \cite{Gregory:2016}. The Zeeman component of the Hamiltonian has only nuclear spin and rotational components
\begin{equation}
H_{\mathrm{Zeeman}}^{\mathrm{RbCs}} = -g_\mathrm{r} \mu_\mathrm{B} \vec{N}\cdot\vec{B}-\sum_{i=\mathrm{Rb,Cs}}g_\mathrm{N}^i\mu_\mathrm{N}\vec{I}_i\cdot\vec{B}.
\label{eq:Zeeman RbcS}
\end{equation}

At zero magnetic field, the states of RbCs are well described by the quantum number $F$, which is the resultant of $N$, $I_\mathrm{Rb}$ and $I_\mathrm{Cs}$. In the ground rotational state ($N=0$), this gives 4 states with $F=2,3,4,5$ separated by multiples of $c_4=19.0(1)~\mathrm{kHz}$\cite{Gregory:2016}. Applying a magnetic field splits these into $(2I_\mathrm{Rb}+1)(2I_\mathrm{Cs}+1)(2N+1)$ separate Zeeman sub-levels, as shown in figure~\ref{fig:RbCs Structure}(a). This gives 32 distinct hyperfine states in $N = 0$, and 96 in $N = 1$, which is significantly more than in CaF. In the limit of high magnetic fields, the rotational and nuclear angular momenta decouple and the states are well represented by $\ket{N,M_N,M_I^{\mathrm{Rb}},M_I^\mathrm{Cs}}$. The measurements reported in this work are performed at the magnetic field of 181.5~G used for STIRAP. This field is not high enough to decouple $N$ and $I$ nor low enough for $F$ to be a good quantum number; the only good quantum number for the angular momentum projection is $M_F=M_N+M_I^\mathrm{Rb}+M_I^\mathrm{Cs}$. 

Understanding the AC Stark effect is vital to implementing internal-state microwave control for polar molecules confined in an optical trap. For RbCs, we have previously measured the scalar and tensor parts of the AC polarisability, $\alpha^{(0)}$ and $\alpha^{(2)}$, for linearly polarised light at a wavelength of 1550\,nm, and investigated the effect of varying the angle between the polarisation of the light and the applied magnetic field~\cite{Gregory:2017}. We showed that coupling between neighbouring hyperfine states manifests in a rich and highly complicated structure with many avoided crossings. In addition, we found that the energy structure is simplest when the polarisation of the light is parallel to the direction of the magnetic field, as this avoids competition between magnetic and electric quantisation axes; this is the case shown in figure~\ref{fig:RbCs Structure}(b) and used in this work.  


\section{Ramsey interferometry}\label{Sec:Ramsey}
\begin{figure}[t!]
\includegraphics[width=\textwidth]{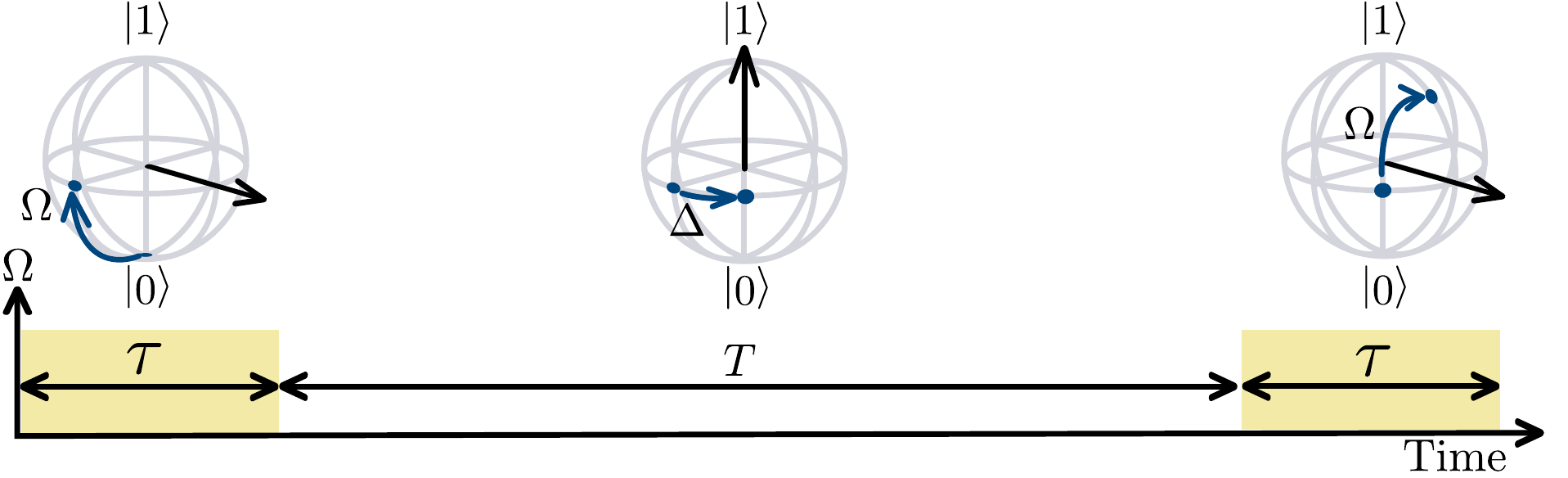}
\caption{The Ramsey interferometry sequence used in our experiments and described in the main text. The time for a $\pi/2$ pulse is $\tau$ and the time between pulses is $T$. }
\label{fig:Ramsey Cartoon}
\end{figure}

Precise measurements of transition frequencies require long interrogation times, which are most readily achieved using trapped samples. Ramsey interferometry is commonly employed to reach the highest precision in metrology and measurements of fundamental constants \cite{Hudson2011,Truppe:2013}, and is also the foundation of the coherent control needed for quantum simulation~\cite{Hazzard:2013}. The method has been used to demonstrate long coherence times between two hyperfine states in the same rotational level in fermionic NaK~\cite{Park2017}. When used to produce a superposition of two rotational states, an oscillating electric dipole moment is induced, introducing dipole-dipole interactions. This technique has been used to observe long-range dipole-dipole interactions between fermionic KRb molecules in an optical lattice \cite{Yan2013}. These ground-breaking results demonstrate the power of these techniques, and refining their use is crucial to advancing the field.

In the Ramsey experiments detailed below, we apply microwave fields tuned close to resonance with a transition between selected states in the ground and first-excited rotational levels of the molecules, here denoted $\ket{0}$ and $\ket{1}$ respectively. Figure~\ref{fig:Ramsey Cartoon} illustrates the sequence. In each experiment, the molecules are first prepared in $\ket{0}$, and then a $\pi/2$ microwave pulse of duration $\tau$ and detuning $\Delta$ creates a coherent superposition of $\ket{0}$ and $\ket{1}$. This state is allowed to evolve for a time $T$ before a second identical $\pi/2$ pulse completes the sequence. In the absence of experimental imperfections, the density operator at the end of the sequence is
\begin{equation}
\rho^{\rm final} = U_{\pi/2} \cdot U_{\mathrm{free}} \cdot U_{\pi/2} \cdot \rho^{\rm initial} \cdot U_{\pi/2}^\dag \cdot U_{\mathrm{free}}^\dag \cdot U_{\pi/2}^\dag,
\label{eq:RamseyRho}
\end{equation}
with $\rho^{\rm initial}=\ket{0}\bra{0}$ the initial density operator. The propagators for the $\pi/2$ pulse, $U_{\pi/2}$, and for the free evolution period, $U_{\mathrm{free}}$, are given by
\begin{align}
U_{\pi/2} &= \cos\left(\frac{X}{2}\right) \hat{\sigma}_0
- i \frac{(\pi/2)}{X}\sin\left(\frac{X}{2}\right) \hat{\sigma}_2
+ i \frac{\tau\Delta}{X}\sin\left(\frac{X}{2}\right) \hat{\sigma}_3,\\
U_{\mathrm{free}} &= \cos\left(\frac{T\Delta}{2}\right) \hat{\sigma}_0
+ i \sin\left(\frac{T\Delta}{2}\right) \hat{\sigma}_3,
\end{align}
where $\hat{\sigma}_i$ ($i \in \{0, 1, 2, 3\}$) are the usual Pauli operators in the Hilbert space spanned by $\ket{0}$ and $\ket{1}$ and $X = \sqrt{(\pi/2)^2 + \Delta^2\tau^2}$. The populations in $\ket{0}$ and $\ket{1}$ at the end of the sequence are $P_0 = \rho_{00}^{\rm final}$ and $P_1 = \rho_{11}^{\rm final}$. 

The theory presented here considers only a single molecule. In the experiments described below, we use ensembles of molecules to be able to measure the population in $\ket{0}$ with a good signal-to-noise ratio in a single iteration of the experiment. When the molecules are confined in a trap to allow long interrogation times, dephasing can become an issue due to spatially varying transition-frequency shifts across the ensemble. This problem is common to experiments of many types, but we show that a judicious choice of molecular hyperfine states minimises its impact. 

\subsection{CaF}
Our experiments with CaF begin with a sample of about $3 \times 10^{3}$ molecules in a single quantum state, $\ket{0} = \ket{N=0,F=1,M_F=1}$. The ensemble has a temperature of $T_{\rm mol}=55~\mathrm{\mu K}$ and a spatial standard deviation of $\sigma_0 = 1.4~\mathrm{mm}$. Our methods for preparing this sample are described in detail elsewhere~\cite{Truppe2018,Truppe2017b,Devlin2016,Truppe2017,Williams2017,Williams:2018}, so are summarized only briefly here. We produce a beam of CaF from a cryogenic buffer-gas source~\cite{Truppe2018}, then decelerate this beam to low speed using the radiation pressure of counter-propagating frequency-chirped laser light~\cite{Truppe2017b}, which we call the slowing light. The slowed molecules are captured and cooled in a magneto-optical trap (MOT)~\cite{Truppe2017,Williams2017}, and then transferred into a blue-detuned optical molasses~\cite{Truppe2017}, where they are cooled to a much lower temperature by sub-Doppler processes~\cite{Devlin2016}. An optical pumping step prepares the majority in $\ket{N=1,F=0,M_F=0}$, and these are then transferred to $\ket{0}$ using a resonant microwave $\pi$-pulse in the presence of a 60~mG magnetic field. Molecules remaining in $N=1$ are pushed away by pulsing on the slowing light, leaving a pure sample in $\ket{0}$~\cite{Williams:2018}. We investigate the rotational coherence times for freely-expanding molecules, and for molecules trapped magnetically.

\begin{figure}[t!]
	\centering
	\includegraphics[width=\textwidth]{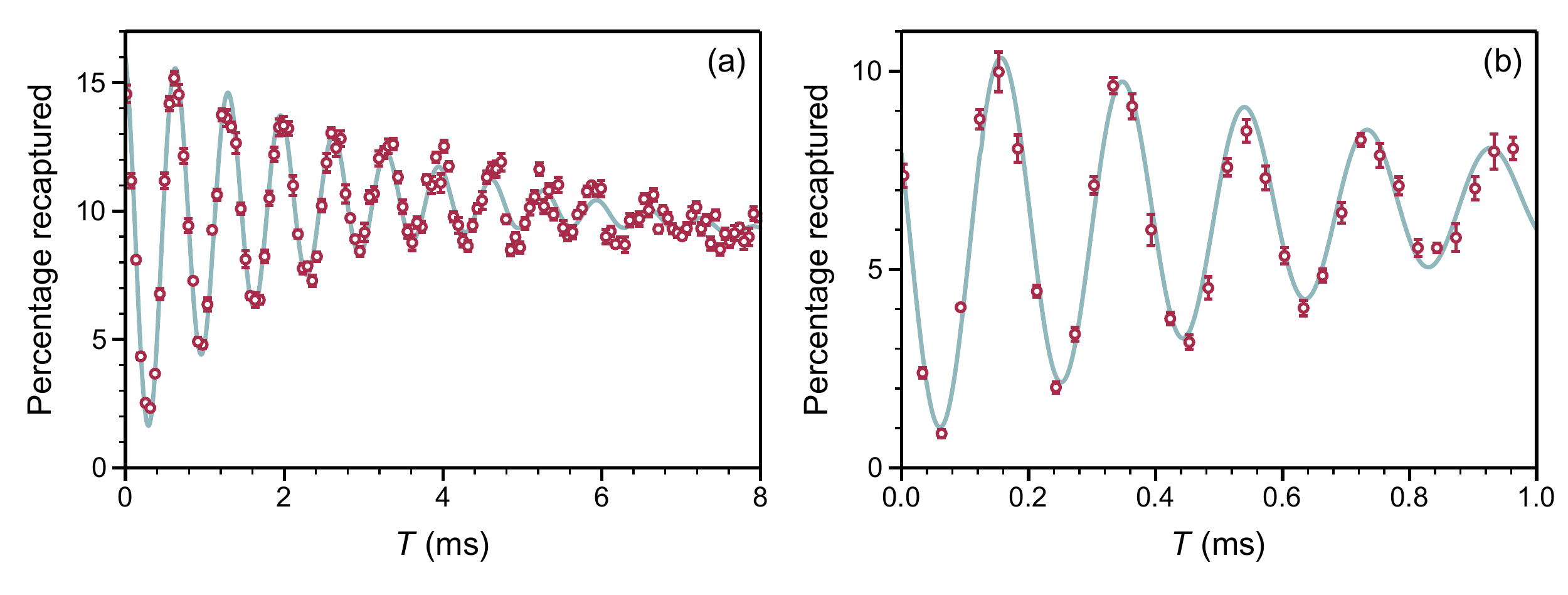}
	\caption[]{Ramsey fringes for CaF molecules prepared in a coherent superposition of states $\ket{0} = \ket{N=0,F=1,M_F=1}$ and $\ket{1}=\ket{N=1,F=2,M_F=2}$. The $\pi/2$ pulses have duration $\tau = 27~\mu$s, and the free evolution time, $T$, is scanned. The plots show the fraction recaptured into the MOT, proportional to the number in $N=1$, as a function of $T$. (a) Molecules in free space. The microwave frequency is 20,553,427.9~kHz, approximately 1.5~kHz above resonance. The main decoherence mechanism is scattering of residual laser light. (b) Molecules confined to a magnetic quadrupole trap with a radial magnetic field gradient of 15~G\,cm$^{-1}$. The microwave frequency is 20,553,431.2~kHz, approximately 5~kHz above resonance. The transition frequency is Zeeman shifted, and the main decoherence mechanism is dephasing due to the distribution of these Zeeman shifts. Points and error bars show the mean and standard error of 9 repeated experiments. Lines are fits to the models described in the text.}
	\label{fig:CaF_Coherence}
\end{figure}

Our Ramsey sequence uses pulses of duration $\tau = 27~\mu$s, tuned close to resonance with the transition from $\ket{0}$ to $\ket{1} = \ket{N=1, F=2, M_F=2}$. At the end of the sequence we measure the number of molecules in $N=1$ by turning on the MOT light and imaging the fluorescence. This number is normalised to the number initially in the MOT, yielding the percentage recaptured. Figure~\ref{fig:CaF_Coherence}(a) shows the percentage recaptured as a function of $T$ for molecules that are freely expanding. Here, we have chosen $\Delta/(2\pi) \approx 1.5~\mathrm{kHz}$, which is small compared to the Rabi frequency. The data show the beat note between the oscillations of the molecule and the microwave source. The coherence time, defined as the $1/e$ decay time of the oscillations, is 2.48(4)~ms. 

To model these results, we introduce two experimental imperfections to the model in equation~\eref{eq:RamseyRho}. The first imperfection concerns leaked light. Although the slowing light is extinguished using an acousto-optic modulator, a small fraction of this light leaks through. It excites molecules that are in state $\ket{1}$, and the excited molecules then decay back to one of the levels of $N=1$. This effect limits the coherence time and pulls the steady-state population in $N=1$ above 50\%. The scattering rate is low enough that we can neglect events occurring during the $\pi/2$ pulses, and concentrate on the free evolution period. We divide the molecules into two groups. The first contains those that have not scattered any photons. The density operator for this group, $\rho^{(1)}$, has matrix elements evolving as $\rho^{(1)}_{11}(T)=\rho^{\pi/2}_{11} e^{-\gamma_{\rm sc} T}$, $\rho^{(1)}_{10}(T) = \rho^{\pi/2}_{10} e^{-\gamma_{\rm sc} T/2 + i\Delta T}$, $\rho^{(1)}_{00}(T)=\rho^{\pi/2}_{00}$, where $\gamma_{\rm sc}$ is the scattering rate and $\rho^{\pi/2} = U_{\pi/2}\cdot\rho^{\rm initial}\cdot U_{\pi/2}^\dag$ is the density operator immediately after the first $\pi/2$ pulse. The second group contains the fraction of molecules that have scattered a photon,
\begin{equation}
f_{\rm scat} = \rho^{\pi/2}_{11} (1-e^{-\gamma_{\rm sc} T}).
\end{equation}
Of these molecules, a fraction $b_r$ are in $\ket{1}$ with density operator $\rho^{(2)} = \ket{1}\bra{1}$ and are affected by the final $\pi/2$ pulse, while the remainder are in other levels of $N=1$ and are unaffected by this pulse. Here, $b_{r}=1/3$ is the probability that a molecule in $\ket{1}$ decays back to $\ket{1}$ after excitation by the slowing light. We can neglect multiple scattering events that further re-distribute the $N=1$ population, since they are rare. The density operator at the end of the Ramsey sequence is now given by
\begin{equation}
\rho'^{\rm \> final} = U_{\pi/2} \cdot \left[\rho^{(1)}(T) + b_r f_{\rm scat} \rho^{(2)}\right]\cdot U_{\pi/2}^{\dagger},
\label{Eq:LeakRho}
\end{equation}
where $U_{\pi/2}$ is the propagator for a $\pi/2$ pulse. The final population measured in $N=1$ becomes
\begin{equation}
P_{1}'(\Delta, \tau, T)  = \rho'^{\rm \> final}_{11} + (1-b_r)f_{\rm scat}.
\label{Eq:leakP}
\end{equation}

The second imperfection is a reduction in $P_{1}'$ with increasing $T$ due to the free expansion of the cloud. We assume there is a cut-off radius, $R$, beyond which the molecules are not detected, and that the size of the cloud expands as $\sigma^2 = \sigma_0^2 + k_{\rm B} T_{\rm mol} T^2/m$. Here, $m$ is the mass of a CaF molecule, and $\sigma_0$ and $T_{\rm mol}$ are fixed at the values given above. The proportion of the total molecules detected is
\begin{equation}
\beta(\zeta) = \mathrm{erf}(\zeta)-\sqrt{4/\pi}\zeta e^{-\zeta^2},
\end{equation}
where $\zeta^2 = R^{2}/(2\sigma^2)$. This modifies the expression for the $N=1$ population to
\begin{equation}
P_{1}''(\Delta, \tau, T)  = \frac{\beta(\zeta)}{\beta(\zeta_0)} P_{1}'(\Delta, \tau, T),
\end{equation}
where $\zeta_0^2 = R^{2}/(2\sigma_0^2)$.

The solid line in figure~\ref{fig:CaF_Coherence}(a) shows a fit to the model $A P_{1}''$ with the amplitude $A$, the detuning $\Delta$, the radius $R$, and the scattering rate $\gamma_{\rm sc}$ as free parameters. The fit has a reduced chi-squared $\chi_{\rm red}^2=7.0$, and gives $\Delta/(2\pi)=1.5053(8)$~kHz, $R=2.54(6)~\mathrm{mm}$ and $\gamma_{\rm sc}=890(10)~\mathrm{s^{-1}}$. This value of $R$ is surprisingly small, suggesting that our simple model may not fully describe the loss mechanism. We note that the statistical uncertainty in determining the transition frequency is less than 1~Hz. Correcting for a frequency shift of $-6(4)$~Hz due to the application of a 60~mG uniform magnetic field, we obtain a preliminary field-free transition frequency of $f_0 = 20,553,426,401(4)$~Hz. We emphasize that we have not yet studied the systematic shifts and uncertainties. The uncertainty in the Zeeman shift can be reduced well below 1~Hz by measuring the frequency at a few values of applied field and extrapolating to zero. The applied field is easily measured to mG accuracy using the Zeeman splitting of the microwave transition. The uncertainty in the first-order Doppler shift can be reduced below 1~Hz by tracking the movement of the cloud with 100~$\mu$m accuracy over a 10~ms period.

\begin{figure}[t!]
	\centering
	\includegraphics[width=\textwidth]{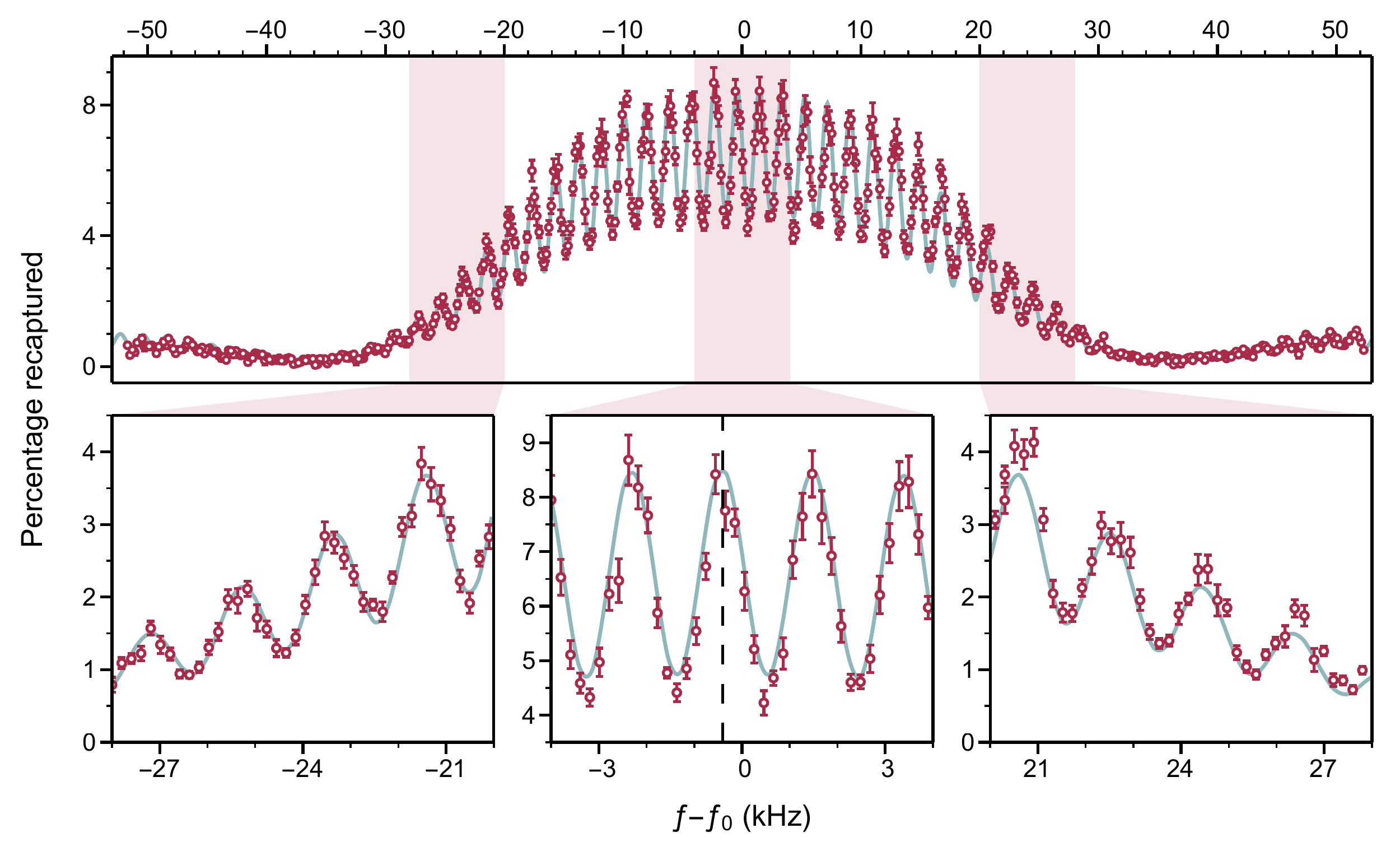}
	\caption[]{Ramsey data for CaF molecules prepared in a coherent superposition of states $\ket{0} = \ket{N=0,F=1,M_F=1}$ and $\ket{1}=\ket{N=1,F=2,M_F=2}$. Molecules are confined to a magnetic quadrupole trap with a radial magnetic field gradient of 15~G\,cm$^{-1}$. The $\pi/2$ pulses have duration $\tau = 27~\mu$s and the free evolution time is $T=493~\mu$s. The plots show the percentage recaptured into the MOT, proportional to the number in $N=1$, as a function of microwave frequency, $f$, relative to the transition frequency measured in free space, $f_0$.
The lower row shows magnified regions of the data. The vertical dashed line in the central plot indicates the position of the central fringe; the transition frequency in the magnetic trap is shifted by $-404(1)$~Hz from $f_0$. Points and error bars show the mean and standard error of 9 repeated experiments. The line is a fit to the model described in the text.} 
	\label{fig:CaF_Lineshape}
\end{figure}

Figure~\ref{fig:CaF_Coherence}(b) shows the same experiment for molecules confined in a quadrupole magnetic trap~\cite{Williams:2018} with a radial magnetic field gradient $A_{\rho}=15$~G\,cm$^{-1}$. All other parameters are the same as above but with a larger detuning of $\Delta/(2\pi) \approx 5$~kHz. In the trap, the Zeeman shift of the transition frequency depends on position, so molecules at different positions fall out of phase, causing decoherence. The Zeeman shifts of states $\ket{0}$ and $\ket{1}$ are almost identical, so long coherence times are possible in the magnetic trap. Nevertheless, the observed coherence time of 0.61(3)~ms is considerably shorter than in free space because of the residual difference between the magnetic moments of the two states, $\mu_{\ket{1}}$ and $\mu_{\ket{0}}$. This difference also shifts the transition frequency measured in the trap. Figure~\ref{fig:CaF_Lineshape} shows another example of data taken using trapped molecules. Here, we fix $T=493~\mu$s, and scan $\Delta$. We observe the standard Ramsey lineshape, but with the contrast reduced, primarily due to the dephasing arising from the distribution of Zeeman shifts. 

To find the distribution of Zeeman shifts, we assume a Gaussian distribution of stationary molecules with standard deviations $\sigma_{\rho}$ and $\sigma_{z}$ in the radial and axial directions. The coherence time observed in figure~\ref{fig:CaF_Coherence}(b) is about 50 times shorter than the typical oscillation period of a molecule in the trap, so the approximation of stationary molecules is a good one. We express the Zeeman shift of the transition as $\delta_{Z} = \eta r$, where $r^2=\rho^2+4z^2$ and $\eta = (\mu_{\ket{1}}-\mu_{\ket{0}}) A_{\rho}/\hbar$. Using the standard method of transforming the variables of a probability density function (pdf), we transform the known pdf of position into a pdf for $\delta_{Z}$. The result is
\begin{equation}
f(\delta_{Z}) = \frac{1}{\eta^2 \sigma' \sigma_{\rho}} e^{-\delta_{Z}^2/(2\eta^2\sigma_{\rho}^2)}\delta_Z\,\mathrm{erfi}\left(\frac{\delta_{Z}\sigma'}{2\sqrt{2}\eta\sigma_{z}\sigma_{\rho}}\right),
\label{Eq:DetuningDist}
\end{equation}
where $\sigma'=\sqrt{4\sigma_z^2-\sigma_{\rho}^2}$, and $\mathrm{erfi}(z) = \mathrm{erf}(i z)/i$. The additional decoherence due to photon scattering is still present, just as in free space, but the loss due to cloud expansion is not present since the molecules are trapped. Therefore, we start with equation~\eref{Eq:leakP}, make the replacement $\Delta \rightarrow \Delta + \delta_{Z}$, and then integrate over the distribution of $\delta_{Z}$ to yield a final expression for the population remaining in $N=1$ after the Ramsey sequence,
\begin{equation}
P_{1}'''(\Delta, \tau, T)  = \int_0^{\pm \infty} P_{1}'(\Delta +\delta_{Z}, \tau, T) f(\delta_{Z})\,d\delta_{Z},
\end{equation}
where the sign of the upper integration limit is the same as the sign of $\mu_{\ket{1}}-\mu_{\ket{0}}$.

We attempt a simultaneous fit of the two datasets shown in figures \ref{fig:CaF_Coherence}(b) and \ref{fig:CaF_Lineshape} to the model $y_{0} + A P_{1}'''$ where $y_{0}$ is a background and $A$ is an amplitude. In this fit, we fix $\tau= 27~\mathrm{\mu s}$, the widths of the trapped distribution to their measured values, $\sigma_{z} = 1.44$~mm and $\sigma_{\rho}=1.37$~mm, and the transition frequency to that measured in free space above. For the data in figure \ref{fig:CaF_Lineshape}(b), we also fix $T=493~\mu$s. We allow separate values of $y_{0}$, $A$ and $\gamma_{\rm sc}$ for each dataset, but single values of the central frequency and of $\eta$ that are common to both sets. The lines in figures \ref{fig:CaF_Coherence}(b) and \ref{fig:CaF_Lineshape} show the results of this simultaneous fit.  The model fits well to both sets of data ($\chi_{\rm red}^2=2.7$), finds values of $\gamma_{\rm sc}$ similar to the one found above, and gives $\eta=8.54(7)\times 10^5~\mathrm{m^{-1}~s^{-1}}$. This value of $\eta$ is determined primarily from the shift in the transition frequency between the free-space measurement and the measurements in the trap. It is sensitive to a possible systematic shift between the position of the cloud loaded into the trap, and the magnetic minimum of the trap. If, instead, we allow the central frequency to float, the only information about $\eta$ comes from the coherence time. In this case, the fit converges on a value of $\eta$ that is 70\% higher.  We take the former fit to be most reliable and use the difference between the two fits as a measure of the uncertainty in $\eta$. We thus obtain a difference in magnetic moments of $\mu_{\ket{1}}-\mu_{\ket{0}}=-7(5) \times 10^{-5}\,\mu_{\rm B}$.

The coherence time in free space is limited in these experiments by scattering of residual laser light. This can be eliminated in future using a fast mechanical shutter. On a longer timescale, the free expansion of the cloud becomes comparable to the wavelength of the microwaves. This limits the coherence time because the change in position between the $\pi/2$ pulses imparts a phase change that is different for each molecule. 

The coherence time in the magnetic trap is limited by the dependence of the transition frequency on magnetic field, which varies across the sample. We have chosen a transition for which the terms with coefficients $g_S$ and $g_{\rm N}^{\rm F}$ in equation~\eref{Eq:HamZeeman} cancel exactly, but there is still a residual difference between the magnetic moments of the two states arising from the terms with coefficients $g_l$ and $g_\mathrm{r}$. As seen above, this difference is difficult to measure precisely, but it can be calculated. We have carried out calculations of $g_\mathrm{r}$ at the Hartree-Fock level, using the DALTON electronic structure package \cite{DALTON,daltonpaper} and cc-pVTZ basis sets \cite{basisF,basisCa}. This gives $g_{\rm r} = -5.15 \times 10^{-5}$ at the CaF equilibrium bond length of 1.95~\AA. We have previously calculated values of $g_{l}$ (sometimes known as $\Delta g_{\perp}$) for a number of $^{2}\Sigma$ molecules~\cite{Aldegunde:2018}. For CaF, our calculation gives $g_{l}=-1.80 \times 10^{-3}$.  Using equation~\eref{Eq:HamZeeman}, and these values for $g_l$ and $g_{\rm r}$, we find that $\mu_{\ket{1}}-\mu_{\ket{0}} = (g_l/15 - g_\mathrm{r})\mu_{\rm B} = -6.85 \times 10^{-5}\mu_{\rm B}$. This is consistent with our measured value. It implies a coherence time of 1.25~ms in the absence of residual laser light. 

\begin{table}
\caption{\label{Tab:MagSensitivity} Difference in magnetic moments, $\Delta \mu = \mu_{\ket{N+1}}-\mu_{\ket{N}}$, between states $\ket{N,F,M_F}=\ket{N,N+1,N+1}$ and $\ket{N+1,N+2,N+2}$ for various $N$. Column 2 is the expression for this difference, column 3 is its value obtained from our calculated $g_l$ and $g_{\rm r}$, and column 4 is the associated coherence time, $\tau_{\rm c}$.}
\begin{indented}	
\item[]\begin{tabular}{@{}cccc}
\br
& \centre{2}{$\Delta \mu$ ($\mu_{\rm B}$)} & \\
\ns
& \crule{2} & \\
$N$ &Expression &Value & $\tau_{\rm c}$ (ms)\\
\mr
0& $g_l/15 - g_{\rm r}$& $-6.85\times 10^{-5}$ & 1.25\\
1& $g_l/35 - g_{\rm r}$& $7\times 10^{-8}$ & 1200\\
2& $g_l/63 - g_{\rm r}$& $2.29\times 10^{-5}$ & 3.74\\
3& $g_l/99 - g_{\rm r}$& $3.33\times 10^{-5}$ & 2.57\\
\br
\end{tabular}
\end{indented}
\end{table}

It is interesting to work out the magnetic sensitivity of other rotational transitions of the type $\ket{N,F=N+1,M_F=F} -\ket{N'=N+1,F'=N'+1,M_F'=F'}$, where only $g_l$ and $g_{\rm r}$ contribute to the Zeeman shift of the transition. Table \ref{Tab:MagSensitivity} shows the predicted difference in magnetic moments between these two states, $\Delta \mu$, for a range of $N$. The table also shows the expected coherence time in a magnetic trap for these transitions, for the experimental conditions presented above and assuming no other decoherence mechanism. Remarkably, $\Delta \mu$ changes sign as $N$ increases and is very close to zero for $N=1$, where the contributions from $g_l$ and $g_{\rm r}$ cancel almost exactly. The exactness of the cancellation is somewhat coincidental, given the $\sim 10$\% accuracy of the calculations. Taking this into account, we still expect $\Delta \mu < 10^{-5}\mu_{\rm B}$ for the $\ket{1,2,2}-\ket{2,2,3}$ transition, and an associated coherence time $t_{\rm c} \gtrsim 10$~ms. For all the transitions, the coherence time could be extended by using a flat-bottomed trap. It may also be possible to tune $\Delta \mu$ even closer to zero by mixing in small fractions of other states.

\subsection{RbCs}
\begin{figure}[t!]
	\centering
	\includegraphics[width=\textwidth]{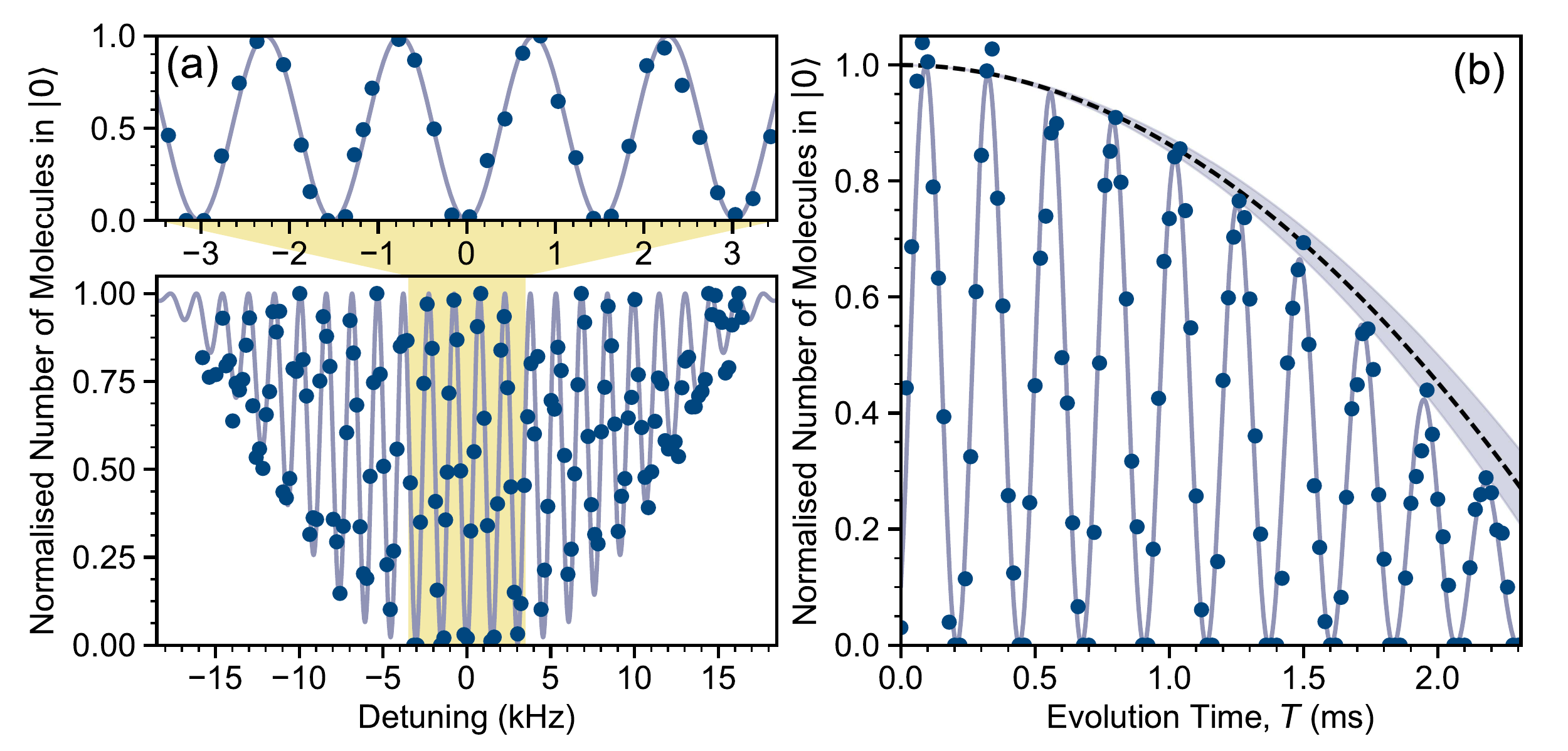}
	\caption{The Ramsey method in free space for RbCs molecules, using a superposition of $\ket{0}=\ket{N=0,M_F=+5}$ and $\ket{1}=\ket{N=1,M_F=+6}$. 
    (a) The $\pi/2$ pulses have duration $\tau =47.8~\mathrm{\mu s}$ and the free evolution time is $T = 600~\mathrm{\mu s}$. The plot shows the normalised number of molecules in $\ket{0}$ as a function of the detuning of the microwave field from resonance. The solid line is a fit ($\chi_{\rm red}^2  = 5.3$) to $P_0=\rho_{00}$ where $\rho$ is defined in equation~\eref{eq:RamseyRho} and yields a transition frequency of $980.385\,569(8)~\mathrm{MHz}$.
    (b) For a fixed microwave frequency of 980.390~MHz, we vary the free evolution time, $T$. We observe Ramsey fringes with a negligible loss of contrast over the 2.3~ms interrogation time. However, we observe a reduction in the molecule number with time as the cloud of molecules falls and expands out of the detection volume (set by the focus of the STIRAP beams). The dashed line shows the result of an independent measurement of this effect with an uncertainty indicated by the shaded region. The solid line is a fit ($\chi_{\rm red}^2  = 1.9$) to $(1-(T/T_0)^2)\times P_0$. In both panels each point represents the result of a single experimental run. 
    }
	\label{fig:RbCs Frequency Ramsey}
\end{figure}
Our experiments on RbCs use samples of up to $4\times 10^3$ ground-state molecules confined in a purely optical trap at a temperature of $1.5~\rm \mu K$ and with a peak density of $2\times10^{11}~\mathrm{cm^{-3}}$. The molecules are created from a mixture of ultracold Rb and Cs atoms using magnetoassociation on an interspecies Feshbach resonance followed by optical transfer by Stimulated Raman Adiabatic Passage (STIRAP). Full details of the steps involved in molecule creation are reported elsewhere~\cite{Jenkin:2011,McCarron:2011,Koppinger:2014,Molony:2014,Gregory:2015,Molony:2016,Molony:2016b}. The STIRAP transfer is performed with hyperfine state resolution, such that the molecules are prepared initially in $\ket{N=0,M_F=+5}$. This is the lowest hyperfine/Zeeman sublevel of the rovibrational ground state at the magnetic field of 181.5~G used in the experiment (see figure~\ref{fig:RbCs Structure}). The molecules are detected by reversing the association sequence and using standard atomic absorption imaging of both Rb and Cs. Due to the state-selective nature of the STIRAP process, our detection is sensitive only to molecules in the initial $\ket{N=0,M_F=+5}$ hyperfine level. 

We begin our investigation of Ramsey interferometry using RbCs by again considering the simplest case of molecules in free space. We prepare a superposition of the two spin-stretched states, $\ket{0} = \ket{N=0,M_F=+5}$ and $\ket{1} = \ket{N=1,M_F=+6}$, as the associated transition is the strongest available between the two rotational levels. In figure~\ref{fig:RbCs Frequency Ramsey}(a), we fix the free evolution time to $T=600~\mathrm{\mu s}$ and vary the detuning of the microwave field from resonance. This yields high-contrast Ramsey fringes spaced by $\approx 1.5~\mathrm{kHz}$ and modulated by an envelope function with a width set by $1 / (2\tau)$. We fit the model for the ground-state population, $P_0$, defined by equation~\eref{eq:RamseyRho} to the results and determine the line centre to be $980.385\,569(8)~\mathrm{MHz}$. Here, the quoted 8~Hz uncertainty is purely statistical. We have not investigated potential systematic shifts and uncertainties. 

To explore the limits of the free-space measurements, we fix the applied microwave frequency to 980.390~MHz (corresponding to a detuning of $\sim5$~kHz) and vary the free evolution time~$T$ as shown in figure~\ref{fig:RbCs Frequency Ramsey}(b). We increase the microwave power such that the $\pi/2$ pulse duration is $\tau = 19~\mathrm{\mu s}$; the detuning is therefore small compared to the Rabi frequency. We observe no loss of fringe contrast over the period of the measurement, indicating a coherence time $>2.5$~ms. However, we observe a reduction in the molecule number with time as the cloud of molecules falls and expands out of the detection volume set by the focus of the STIRAP beams. Although the STIRAP beams have a waist of $\sim 30~\mu$m, the STIRAP transfer efficiency depends sensitively on the intensity and is optimised only near the focus. We have measured this effect independently; the result is indicated by the dashed line in figure~\ref{fig:RbCs Frequency Ramsey}(b). We therefore fit $(1-(T/T_0)^2)\times P_0$ to this measurement, with $T_0= 2.57(4)~\rm ms$. We extract a transition frequency of $980.385\,698(3)~\mathrm{MHz}$, where the quoted 3~Hz uncertainty is again purely statistical. We note that this result is 129(9)~Hz greater than the transition frequency obtained from the measurement in figure~\ref{fig:RbCs Frequency Ramsey}(a). As the microwave source was referenced to an external 10~MHz GPS reference, we believe that the difference stems from a difference in the conditions of the experiment on the separate days that the measurements were performed. The Zeeman shift of the transition is $-4.8$~Hz\,G$^{-1}$~\cite{Gregory:2016}, indicating that the difference is not attributable to a change in the magnetic field (which is typically $<50$\,mG day-to-day). The transition is however sensitive to stray electric fields, which can be present in the UHV glass cell \cite{Molony:2014}; a DC Stark shift of the transition by 129~Hz requires a DC electric field of only 1.2~V~cm$^{-1}$. Alternatively, the difference may result from a more subtle systematic effect such as coupling to nearby hyperfine states; the Fourier width of the microwave pulses is similar to the spacing between neighbouring hyperfine states and different pulse durations were used for the two measurements. These systematic shifts and their uncertainties will be investigated in future work.

\begin{figure}[t!]
	\centering
	\includegraphics[width=\textwidth]{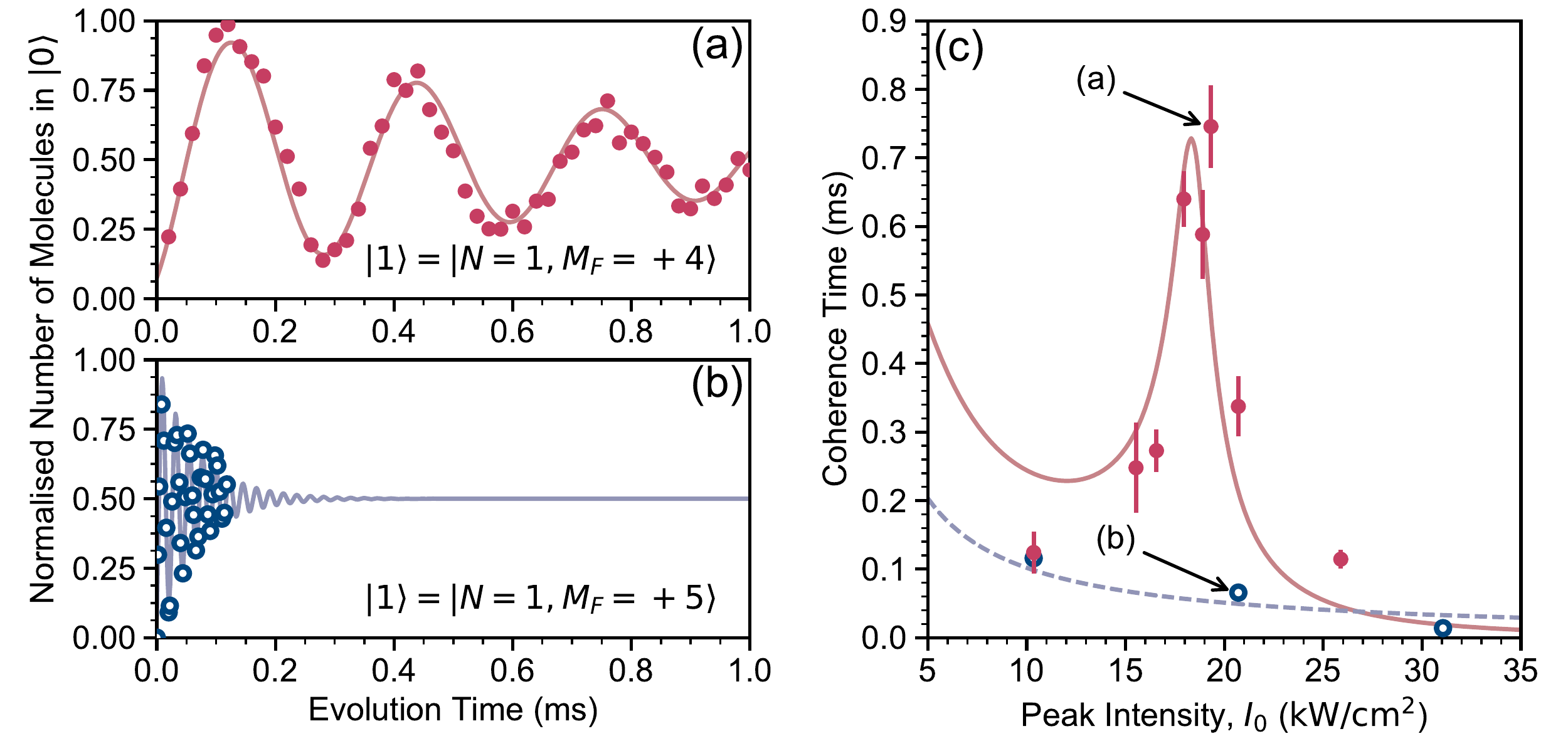}
	\caption{
    Ramsey measurements using RbCs molecules confined in an optical trap.   
    (a) Long-lived Ramsey fringes in a trap with a peak intensity of $I_0=19.3~\mathrm{kW\,cm^{-2}}$ using a superposition of $\ket{0}=\ket{N=0,M_F=+5}$ and $\ket{1} =\ket{N=1,M_F=+4}$. 
    (b) Rapid dephasing of the Ramsey fringes in a trap with a peak intensity of $I_0=20.7~\mathrm{kW\,cm^{-2}}$ using a superposition of $\ket{0}=\ket{N=0,M_F=+5}$ and $\ket{1} =\ket{N=1,M_F=+5}$. In both (a) and (b) each point represents the result of a single experimental run and the solid line is a damped sine-function fit  ($\chi_{\rm red}^2  =0.2,0.4$ respectively)  to the data used to extract a coherence time.
   (c) The coherence time as a function of the peak intensity of the trap for superpositions of $\ket{0}=\ket{N=0,M_F=+5}$ with $\ket{1}=\ket{N=1,M_F=+4}$ (filled red points) and $\ket{1}=\ket{N=1,M_F=+5}$ (open blue points). The lines represent fits of the simple model described in the text and equation \eref{eq:RbCs Coherence Fit}, showing qualitative agreement with our expectation that the coherence is maximised when the differential AC Stark shift across the sample is minimised.
   }
	\label{fig:RbCs Coherence}
\end{figure}

It will ultimately be desirable to interrogate molecules confined in an optical lattice, where longer evolution times are possible and interesting many-body effects may be present. Here we extend our Ramsey measurements to molecules confined in a simple optical trap, in order to determine its impact on the observed coherence time. We achieve this by recapturing the molecules in an optical trap after the STIRAP transfer to the ground state. The trap consists of two linearly polarised beams with $\lambda=1550$~nm and waists of $w_{01}=80$~$\mu$m and $w_{02}=98$~$\mu$m, crossing at an angle of $27^{\circ}$ in the horizontal plane. We set the polarisation of both beams to be along the direction of the magnetic field with an uncertainty of $<3^{\circ}$. The peak intensity of the trap light prior to STIRAP is $37.3~\rm{kW\,cm}^{-2}$ for all the measurements. We vary the intensity of the trap used to recapture the molecules, thereby exploring the effect on the Ramsey sequence of different AC Stark shifts, shown in \fref{fig:RbCs Structure}~(b).  At the same time, the molecules experience a different trap potential depending on the intensity of the trap light~\cite{Gregory:2017}. Creating a trap deep enough to prevent evaporation of the molecules requires a peak intensity $>20~\mathrm{kW\,cm^{-2}}$, and matching the potential to that experienced by the Feshbach molecules requires a peak intensity of $43~\mathrm{kW\,cm^{-2}}$.

In principle, the trap extends the time over which the molecules can be interrogated. However, molecules in different parts of the trap experience different intensities, resulting in a spatially varying AC Stark shift of the microwave transition. The initial distribution of the ground-state molecules reflects that of the Feshbach molecules, as the molecules move a negligible distance during the $50~\mu \rm s$ that the trap is turned off for STIRAP. The distribution is Gaussian, with standard deviations of $\sigma_{z} = 6.6~\mu$m in the vertical direction and $\sigma_\mathrm{axial} = 24~\mu$m in the axial direction. Due to gravitational sag, the centre of the distribution is $z_{0} = 8.1~\mu$m below the position of peak intensity. Under these conditions, the variation of intensity across the cloud is dominated by the vertical direction and we estimate the $2\sigma$ intensity difference to be
\begin{equation}
\Delta I \approx \frac{8 z_{0} \sigma_{z}}{w_{0}^{2}} I_{0} \approx 0.04 I_{0}.
\label{eq: intensity distribution}
\end{equation}
Crucially, this depends on the peak intensity $I_0$, indicating that the spread of intensities is greater for deeper traps.
For simplicity, we assume that the molecular distribution and the associated intensity variation remains constant during the measurements. Typical trap oscillation periods are $\sim 5$\,ms, such that this approximation is valid for measurements performed in under $\sim 0.5$\,ms. For longer times, the intensity variation will be greater than the above estimate, as the molecular cloud will fall and expand, since all the intensities investigated are below the $43~\mathrm{kW\,cm^{-2}}$ needed to match the trap potential. Although this effect is undoubtedly important for some of our measurements, this approximate model gives sufficient insight for the present work.




To measure the effect of the trap light on the coherence time, we perform Ramsey measurements by varying the free evolution time $T$. We use different hyperfine levels of $N=1$ compared to the free space measurement. Specifically, we investigate superpositions of $\ket{0} = \ket{N=0,M_F=+5}$ with either $\ket{1} = \ket{N=1,M_F=+4}$ or $\ket{1} =\ket{N=1,M_F=+5}$. We choose these states as, unlike $\ket{N=1,M_F=+6}$, their transitions are well separated in frequency from other nearby transitions, as shown in figure~\ref{fig:RbCs Structure}(b), minimising the possibility of off-resonant coupling. To perform each measurement, we turn on the trap light to recapture the molecules in $\ket{N=0, M_{F}=+5}$. We then wait $400~\mu\rm s$ before performing the Ramsey sequence. This ensures that the intensity of the light is stable prior to the measurement; the intensity is monitored on a photodiode, and stabilised by an active servo loop with a bandwidth of $\sim 50$\,kHz. For each trap intensity, we first determine the transition frequency and measure the Rabi frequency at zero detuning to define the $\pi/2$ pulse duration, $\tau$. For $\ket{N=1,M_F=+5}$ we typically use $\tau \approx 2.5~\mu\rm s$, whilst for $\ket{N=1,M_F=+4}$ we use $\tau \approx 12~\mu\rm s$. Note that the transition to the $M_F=+4$ state is considerably weaker and its strength varies appreciably with laser intensity. For this transition, we must therefore use a longer $\tau$ and are limited to smaller detunings $\Delta$. Figures~\ref{fig:RbCs Coherence}(a) and (b) show typical results of such measurements for the two different hyperfine levels. We observe Ramsey fringes which decohere with time as the spatially dependent AC Stark shift across the cloud leads to dephasing. To quantify this effect, we fit each dataset to a sine wave with an exponentially decaying amplitude in order to extract a coherence time $\tau_\mathrm{c}$.

The results shown in figures~\ref{fig:RbCs Coherence}(a) and (b) exhibit significantly different coherence times for the two transitions, despite very similar peak trap intensities. This difference stems from the different AC Stark shifts shown in figure~\ref{fig:RbCs Structure}(b). For $\ket{N=1,M_F=+5}$, the AC Stark shift is very close to linear. The spread of transition frequencies across the cloud is then simply $\Delta f = (\mathrm{d} f/\mathrm{d}I)\times \Delta I$ and the associated dephasing time is expected to scale inversely with the peak intensity. For the peak intensity of $20.7~\rm{kW\,cm}^{-2}$ shown in figure~\ref{fig:RbCs Coherence}(b),  we find that a superposition between $\ket{N=0,M_F=+5}$ and $\ket{N=1,M_F=+5}$ has a coherence time of $66(5)~\mu\rm s$. In contrast, the transition to $\ket{N=1, M_F=+4}$ displays a broad avoided crossing around $16~\rm{kW\,cm}^{-2}$, where $(\mathrm{d} f/\mathrm{d}I)\simeq 0$ and the variation of transition frequency across the cloud is minimised. We observe a significant increase in the coherence time for this state around the avoided crossing, as shown in figure~\ref{fig:RbCs Coherence}(c). The maximum coherence time that we measure for $\ket{N=1, M_F=+4}$ in the trap is 0.75(6) ms, and is approximately an order of magnitude greater than that achieved using $\ket{N=1,M_F=+5}$.

To model the results in figure~\ref{fig:RbCs Coherence}(c) we need accurate knowledge of the AC Stark shift of the transitions. The prediction shown in figure~\ref{fig:RbCs Structure}(b) is for the polarisation of the trap light \textit{exactly} aligned with the direction of the magnetic field. Even small deviations from this condition can lead to significantly different AC Stark shifts, particularly around an avoided crossing~\cite{Gregory:2017}. We therefore use the period of the Ramsey fringes to determine the transition frequency for each intensity, effectively mapping out the AC Stark shift under the conditions of the experiment (i.e. accounting for any small misalignment of the polarisation of the trap beams). For the transition to $\ket{N=1,M_F=+4}$, we fit the measured frequencies to a third-order polynomial constrained to the known zero-intensity transition frequency~\cite{Gregory:2016}. For the transition to $\ket{N=1,M_F=+5}$, we use a simple linear fit.
For a given intensity $I$, we extract the minimum and maximum transition frequencies in the range $I-\Delta I/2 \rightarrow I+\Delta I/2$  to determine the spread of transition frequencies $\Delta f(I)$ responsible for the dephasing of the Ramsey signal. The total coherence time $\tau_\mathrm{c}$ is given by
\begin{equation}
\tau_\mathrm{c} =\left[\left(\frac{1}{T_2}\right)^2 + (2\pi \times \Delta f(I))^2 \right]^{-\frac{1}{2}},
\label{eq:RbCs Coherence Fit}
\end{equation}
where $T_{2}$ is the coherence time from all other sources of decoherence in the experiment.

The solid lines in figure~\ref{fig:RbCs Coherence}(c) show the results of fitting the simple model described by equation~\eref{eq:RbCs Coherence Fit} to the measurements of the coherence time, with $T_2$ and $\Delta I$ as fit parameters. For the transition to $\ket{N=1,M_F=+4}$, we find $T_{2}=0.7(2)$~ms and a spread of intensities equal to 3.4(9)\,\% of the peak intensity. The measurements using $\ket{N=1,M_F=+5}$ are adequately described by the intensity-dependent dephasing term alone and the fit yields a 2(1)\,\% intensity variation. In both cases, the spread in intensity is in reasonable agreement with the simple estimate of equation \eref{eq: intensity distribution}.
The fitted $T_2$ time is considerably shorter than the coherence time observed in free space. This is most likely due to the lack of a full dynamical model of the molecular motion leading to an underestimate of the intensity variation for longer evolution times.  

In future work we plan to eliminate dynamical effects by loading the molecules into a 3D optical lattice, such that each molecule is pinned on a site of the lattice. Under such conditions spin-echo sequences can be employed to combat dephasing~\cite{Yan2013} and we can investigate the limits on the coherence time more thoroughly. We have shown that avoided crossings in the AC Stark shift can be used to extend the coherence time in the trap greatly. A key challenge for the future is to identify the best avoided crossings to use in the optical lattice. The exact structure of hyperfine levels under the effect of the AC Stark shift is dependent on the magnetic field, as well as the intensity and polarisation of the light. This gives a large parameter space in which to optimise the AC Stark shift for future experiments.

\subsection{Summary}
We have demonstrated the use of Ramsey interferometry in two, very different, ultracold molecule experiments using CaF and RbCs. In both cases, we are able to control the internal rotational and hyperfine states to a sufficient degree that high-contrast Ramsey fringes can be observed and used to determine transition frequencies with precisions on the hertz level. To increase the interrogation time available, we confine the molecules using a magnetic trap for CaF and an optical trap for RbCs. Both traps introduce spatially varying energy-level shifts, leading to dephasing of the Ramsey fringes. In the case of CaF, differential Zeeman shifts limited the coherence time to 0.61(3)~ms for molecules in a magnetic trap. In RbCs we find that the choice of hyperfine states is very important and that avoided crossings in the AC Stark shift can be exploited to reduce dephasing and extend the coherence time to 0.75(6)\,ms.  Finally we have provided suggestions that could greatly suppress the dephasing due to spatially varying energy shifts, either through mixing states with different magnetic moments or by tuning the parameters of an optical lattice to engineer broad avoided crossings.  

\section{Outlook and conclusion\label{sec:Outlook}}


Many proposals for quantum simulation require addressing and detection of individual particles in an ordered array. For molecules this is a major experimental challenge. In this concluding section we will briefly discuss two potential experimental realisations of ordered molecular arrays: a quantum gas microscope and an array of optical micro-traps.

\subsection{Building arrays of ultracold molecules for quantum simulation} 

One approach to building arrays of molecules, used with great success for atoms, is to load the molecules onto individual sites of an optical lattice and use high-resolution single-site imaging, as employed in quantum gas microscopes~\cite{Bakr:2009,Sherson:2010,Cheuk:2015}. Direct loading of an optical lattice from a 3D trap requires a high initial density $\sim(\lambda / 2)^{-3}$, where $\lambda/2$ is the lattice spacing. Currently, this approach is therefore restricted to molecules prepared by the association method. Here, the initial atomic gases can be cooled to quantum degeneracy and efficiently loaded into the lattice in a Mott insulator (MI) state~\cite{Jaksch:1998,Greiner:2002} such that two atoms reside in each site prior to association~\cite{Jaksch:2002,Danzl:2010, Moses:2015, Covey:2016, Reichsollner:2017}. This approach greatly increases the efficiency of the magnetoassociation step to close to unity~\cite{Thalhammer:2006, Chotia:2012}, as well as producing molecules in an ordered array. However, in two-species experiments, achieving a high filling fraction of heteronuclear atom pairs is difficult and requires careful control of both the intraspecies and interspecies interactions. Nevertheless, molecular filling fractions of up to $\sim 30$\,\% in 3D lattices have been demonstrated for both ground-state KRb molecules~\cite{Moses:2015} and RbCs Feshbach molecules~\cite{Reichsollner:2017}. At such fillings, coherent many-body effects due to the long-range DDI are predicted to be observable~\cite{Kwasigroch:2017}.
In both these experiments, interspecies Feshbach resonances are exploited to enhance the miscibility of the atomic mixture. However, this can introduce high inelastic losses for one of the species. Fortunately the MI phase can also be used to protect against such inelastic collisions. For example, Cs suffers from high 3-body loss rates at most magnetic fields, but by first loading the atoms into a MI state with one atom per lattice site at a magnetic field with favourable intraspecies interactions, the atoms can be protected against collisions when moving to a magnetic field with favourable interspecies interactions~\cite{Reichsollner:2017}. Further exploiting the advanced control techniques developed for atomic gases, for example by using species-specific optical potentials~\cite{LeBlanc:2007}, it should be possible to create molecules in optical lattices with even higher filling factors. 

Accurate site-resolved detection and addressing of individual molecules in an optical lattice presents additional challenges. In the atomic quantum gas microscopes, a lens with a high numerical aperture (NA) is used to collect scattered photons as the atoms are laser-cooled in a very deep optical lattice~\cite{Bakr:2009,Sherson:2010,Cheuk:2015}. The small depth of field allows site-resolved imaging only of a single plane of atoms, although sequential readout has been employed to study a bilayer system~\cite{Preiss:2015}. It is important to note that atoms are detected only on sites with an odd number of atoms; light-assisted inelastic collisions lead to rapid loss of both atoms on doubly occupied sites. This \textit{parity detection} has led to striking images of the Mott-insulator transition~\cite{Bakr:2010}. For molecules produced by association, detection usually involves reversing the association process and detecting the constituent atoms, although direct absorption imaging has been demonstrated~\cite{Wang:2010}. In a lattice, reversing the association will inevitably lead to two atoms on each site. To implement site-resolved imaging, either the atoms must be separated using species-specific optical potentials~\cite{LeBlanc:2007} or one species must be removed, for example by using a short pulse of blue-detuned light. Addressing of individual molecules can be achieved by focusing light onto a specific lattice site using the same high-NA lens used for imaging. The resulting AC Stark shift, experienced only by the molecule on the chosen site, allows a microwave field to address just a single molecule. The desire to manipulate molecules with electric fields imposes additional constraints on the microscope, as the inclusion of electrodes forces the use of lenses with longer working distances. However, several designs have been proposed to ensure that the electric field is uniform across the lattice whilst keeping the electrode structure sufficiently compact~\cite{Gempel:2016, Bohn:2017}. 

The direct loading of an optical lattice is appealing when associating atoms to form molecules, as the production of high-phase-space density samples of many atomic species is well understood. However, a variety of interesting experiments could be done with smaller arrays using tweezer traps, which may be loaded from lower-density sources. A tweezer trap is a tightly focussed far-detuned laser beam that can confine a single atom. When loaded from a MOT or molasses, the occupancy of the trap is always zero or one, because two atoms in the same trap undergo a fast light-assisted collision involving the laser-cooling light that ejects both~\cite{Schlosser:2001}. Multiple tweezer traps can be loaded stochastically, and then re-arranged to make regular, defect-free 1D, 2D or even 3D arrays~\cite{Miroshnychenko2006, Lee2016, Barredo:2016, Barredo2018}. Exciting the atoms to Rydberg states introduces strong, controllable interactions between them, and the many-body quantum dynamics of a linear array of such atoms has recently been studied~\cite{Bernien2017}. The same techniques could be used to make arrays of ultracold molecules with dipole-dipole interactions. Molecules could be loaded into tweezer traps directly from a MOT or molasses; the phase-space density reached with laser-cooled molecules is already high enough for efficient loading. Alternatively, molecules could be formed by associating pairs of atoms pre-loaded into the same tweezer~\cite{Liu:2018}. These are attractive approaches to quantum simulation and quantum-information processing with molecules. The arrays are easily reconfigured, so various geometries and molecule spacings can be explored. Molecules can be brought together to interact at a fixed spacing for a known time, then separated again to turn off the interaction and read out the final state of each. 

Atoms in tweezer traps have been cooled to the motional ground state by Raman sideband cooling~\cite{Kaufman2012, Thompson2013}. The cooling proceeds by driving a Raman transition from $|i,n\rangle$ to $|j,n-1\rangle$, followed by optical pumping back to $|i,n-1\rangle$. Here $i,j$ are internal states, typically different hyperfine states, and $n$ is the motional quantum number. This continues until the system reaches $|i,0\rangle$ which is dark to both the Raman and optical pumping steps. It is desirable to cool molecules in the same way; for CaF, both the optical pumping and Raman steps can in principle be done using one of the laser-cooling transitions. However, because of the tensor nature of the polarizability, the AC Stark shift is different for each rotational, hyperfine and Zeeman sub-level. The resulting state-dependent trapping potential brings several complications. The trap frequency will, in general, be different for states $|i\rangle$ and $|j\rangle$, which means that the frequency of the Raman transition depends on $n$, whose value is unknown when the cooling begins. The usual theory used to find the relative strengths of transitions between motional states $n$ and $m$, involving an expansion in the Lamb-Dicke parameter, no longer applies because the states $|i,n\rangle$ and $|j,m\rangle$ are not orthogonal. Instead, Franck-Condon factors between these states determine the transition strengths. The spatially varying elliptical polarization of the light around the focus of the tweezer, resulting from the breakdown of the paraxial approximation, adds further complexity~\cite{Thompson2013}. State-dependent potentials can also result in rapid dephasing of the coherences needed for quantum simulation, as explored in section~\ref{Sec:Ramsey}. However, these challenges also bring new opportunities. For example, with state-dependent potentials, efficient sideband cooling to the motional ground state could be done using microwave transitions between rotational states, instead of Raman transitions~\cite{Li2012}. We are currently analysing these issues with the aim of finding the best approach to sideband cooling and the control of coherence for molecules in tweezer traps.

\subsection{Concluding remarks}
Ultracold molecules offer a variety of new possibilities for quantum simulation, but the development of experiments presents a number of key challenges. We have presented our progress towards overcoming these challenges to develop a platform on which to build future quantum simulation experiments. The molecules used in this work, CaF and RbCs, are produced in very different ways, using two of the most successful approaches in the field, laser cooling and magnetoassociation. We have described in detail the hyperfine and rotational structure of each molecule, highlighting their differences and similarities. 
We have demonstrated Ramsey spectroscopy of the rotational transition $N=0 \leftrightarrow N=1$ for free-space and trapped samples in each experiment.
In the case of trapped samples, we find that the main decoherence mechanism is dephasing due to spatially varying Zeeman shifts for magnetically trapped CaF or AC Stark shifts for optically trapped RbCs. For both molecules, we find that the choice of hyperfine states is critical to maximise the coherence time. Future experiments will focus on the loading of molecules into arrays, either by direct loading into optical lattices, or by building smaller arrays of tweezer traps. This will open the door to a number of realisable quantum simulation experiments.

\ack
This work was supported by U.K. Engineering and Physical Sciences Research Council (EPSRC) Grants EP/P01058X/1, EP/P009565/1, EP/P008275/1, EP/M027716/1, EP/K038311/1 and EP/I012044/1.
D.~Jaksch acknowledges support from the European Research Council Synergy Grant Agreement No.\ 319286 Q-MAC.
L.~Caldwell, B.~E.~Sauer and M.~R.~Tarbutt acknowledge support from the European Research Council under the European Union's Seventh Framework Programme (FP7/2007-2013) / ERC grant agreement 320789.
J.~Aldegunde acknowledges funding by the Spanish Ministry of Science and Innovation, Grants No. CTQ2012-37404-C02 and CTQ2015-65033-P, and Consolider Ingenio 2010 CSD2009-00038.
Data and analysis from this work are available at doi:10.15128/r2kp78gg38.

\section*{References}
\bibliography{QSUM_references_used}
\bibliographystyle{iopart-num}
\end{document}